%% file: thinkair.tex
\documentclass[letter]{sig-alternate-10pt}

\pdfoutput=1

\usepackage{url}
\usepackage{booktabs}
\usepackage{multirow}

\usepackage[usenames,dvipsnames]{color}

\newcommand{\one}{({\em i})}
\newcommand{\two}{({\em ii})}
\newcommand{\three}{({\em iii})}
\newcommand{\four}{({\em iv})}

\newcommand{\idle}{\textsc{idle}}
\newcommand{\dedicated}{\textsc{cell\_dedicated}}
\newcommand{\shared}{\textsc{cell\_shared}}

\usepackage{listings}
\definecolor{gray}{gray}{0.4}
\begin{document}

\title{Unleashing the Power of Mobile Cloud Computing using ThinkAir}

\numberofauthors{3}
\author{
\alignauthor Sokol Kosta\\
\affaddr{Deustche Telekom Labs}\\
\affaddr{Berlin, Germany}\\
\email{kosta@di.uniroma1.it}
\alignauthor Andrius Aucinas\\
\affaddr{University of Cambridge}\\
\affaddr{Cambridge, UK}\\
\email{aa535@cam.ac.uk}
\alignauthor Pan Hui\\
\affaddr{Deutsche Telekom Labs}\\
\affaddr{Berlin, Germany}\\
\email{pan.hui@telekom.de}
\and
\alignauthor Richard Mortier\\
\affaddr{University of Nottingham}\\
\affaddr{Nottingham, UK}\\
\email{rmm@cs.nott.ac.uk}
\alignauthor Xinwen Zhang\\
\affaddr{Huawei America Research}\\
\affaddr{Santa Clara, CA, USA}\\
\email{xinwen.zhang@huawei.com}
}

\maketitle

\input{abstract}

\section*{Keywords}

Mobile Cloud Computing, Smartphone, Virtual Machine, Power Consumption, Code Offloading

\input{intro}

\input{related}

\input{arch}

\input{execenv}
\input{appserver}
\input{profilers}
\input{eval}

\input{discuss}
\input{concl}

\bibliographystyle{unsrt}
{\bibliography{thinkair}}

\end{document}

%% file: abstract.tex
\section*{ABSTRACT}

Smartphones have exploded in popularity in recent years, becoming ever
more sophisticated and capable. As a result, developers worldwide are
building increasingly complex applications that require ever
increasing amounts of computational power and energy.
In this paper we propose \emph{ThinkAir}, a framework that makes it simple for
developers to migrate their smartphone applications to the cloud. ThinkAir exploits the concept of smartphone virtualization in the
cloud and provides method level computation of\mbox{}f\mbox{}loading.
Advancing on previous works, it focuses on the elasticity and scalability
of the server side and enhances the power of mobile cloud computing
by parallelizing method execution using multiple Virtual Machine (VM) images.
We evaluate the system using a range of benchmarks starting from simple
micro-benchmarks to more complex applications.
First, we show that the execution time and energy consumption decrease
\emph{two} orders of magnitude for the $N$-queens puzzle and \emph{one} order of magnitude
for a face detection and a virus scan application, using cloud of\mbox{}f\mbox{}loading.
We then show that if a task is parallelizable, the user can request
more than one VM to execute it, and these VMs will be provided dynamically.
In fact, by exploiting parallelization, we achieve a greater reduction on the execution
time and energy consumption for the previous applications.
Finally, we use a memory-hungry image combiner tool
to demonstrate that applications can dynamically request
VMs with more computational power in order to meet their computational requirements.

%% file: intro.tex
\section{INTRODUCTION}
\label{s:intro}

Smartphones are becoming increasingly popular, with current reports
stating that approximately 350,000 new Android devices are being activated
worldwide every day\footnote{\url{http://finance.yahoo.com/news/350000-Google-Android-Devices-twst-1887349177.html?x=0&.v=1}}.
These devices have a wide range of capabilities, typically including
GPS, WiFi, cameras, gigabytes of storage, and gigahertz-speed
processors. As a result, developers are building ever more complex
smartphone applications that support gaming,
navigation, video editing, augmented reality, and speech recognition
which require considerable computational power and energy.
Unfortunately, as the applications become more complex, users must
continually upgrade their hardware to keep pace with the applications'
requirements, and still experience short battery lifetimes with newer hardware.

To address the issues of computational power and short battery lifetimes,
there has been considerable current research.
Prominent among those are the MAUI~\cite{maui} and the \textit{CloneCloud}~\cite{clonecloud:2011} projects.
MAUI provides method level code offloading based on the Microsoft .NET framework.
However, they allocate an individual application server to each application,
which makes the MAUI framework non-scalable to efficiently admitting new applications.
The CloneCloud project~\cite{clonecloud:2011} proposes a neater management framework for mobile
cloud computing than MAUI with respect to scalability,
by cloning the whole OS image of the cellular phone to the cloud.
Their approach is process-based, i.e., tries to extrapolate pieces of the binary
of a given process whose execution on the cloud would make the overall process execution faster.
They determine these parts by the use of an offline pre-processing static analysis of different
running conditions of the process' binary on both the target smart-phone and the cloud.
The output of such analysis is then used to build a data-base of pre-computed partitions
of the binary code that will eventually be used to determine which parts should be migrated on the cloud.
However, this approach is limited to runs whose input/environmental conditions have been considered
in the offline pre-processing.
Furthermore it needs to be booted for every new application build by developers.



In this paper, we propose \textit{ThinkAir}, a new mobile cloud computing framework which takes the best of the two worlds. It mitigates the MAUI's bottleneck of 
having a server application for each application by cloning the whole device's OS on the cloud and release the system from the restrictions of only previously considered applications/inputs/environmental conditions that CloneCloud induces by adopting an online method-level offloading. Moreover, ThinkAir (1) provides an efficient
way to perform on-demand resource allocation, and (2) exploits parallelism
by dynamically creating, resuming, and destroying VMs when needed.
To the best of our knowledge, ours is the first contribution to address the latter two points in mobile clouds.
The problem of on-demand resource allocation is important because of the following scenario:
let us consider a commercial cloud provider serving multiple smartphone users with
commercial grade services. Users may request different computational
power based on their workload and deadline for tasks, and hence the provider has to
dynamically adjust and allocate its resources to satisfy customer expectations.
Existing research works do not provide any mechanism to perform
on-demand resource allocation, which is an absolute necessity given the
variety of applications that can be run on the mobile smartphones,
in addition to the high variance of CPU and memory requirements these applications could demand.
The problem of exploiting parallelism is important because many current applications
require large amounts of processing power, and parallelizing application processing
reduces execution time and energy consumption of these applications by significant
margins when compared to non-parallel executions of the same.

ThinkAir achieves all the above mentioned goals by providing the profilers and
infrastructure to make efficient and effective code migration
possible; library and compiler support to make it easy for
developers to exploit it with minimal modification of existing code; VM manager
and parallel processing module to dynamically create, resume, suspend, and
destroy smartphone VMs as well as automatically split and distribute tasks to
multiple VMs\footnote{As we use VM to clone the image of a smartphone in the
cloud, we use VM and clone interchangeably in the paper.}.

We now continue by positioning ThinkAir with respect to related
work~(\S\ref{s:related}) before outlining the ThinkAir
architecture~(\S\ref{s:architecture}). We then describe the three
main components of ThinkAir in more detail: the execution
environment~(\S\ref{s:execenv}), the application
server~(\S\ref{s:appserver}), and the profilers~(\S\ref{s:profilers}).
Finally, we evaluate the performance of
ThinkAir~(\S\ref{s:eval}), discuss design limits and future plans~(\S\ref{s:discuss}), and conclude the paper~(\S\ref{s:concl}).

%% file: related.tex
\section{RELATED WORK}
\label{s:related}

Mobile cloud computing has become a hot topic in the community in
recent years.  The basic idea of dynamically switching between
(constrained) local and (plentiful) remote resources, often referred
as cyber-foraging,
has shed light on many research work~\cite{Aura,Balan03-tactics,Spectra,Odyssey,Porras-foraging,dimorphic}.  These approaches augment the capability of
resource-constrained devices by offloading computing tasks to nearby
computing resources, or \emph{surrogates}.  ThinkAir takes insights and inspirations from these previous systems, and shifts the focus from alleviating memory constraints and provide evaluation on
hardware of the time, typically laptops, to more modern smartphones. Furthermore, it enhances computation performance by exploiting parallelism with multiple VM creation on elastic cloud resources and provides a convenient VM management framework for different QoS expectation~\cite{Berkeley-cloud-view}.

Several approaches have been proposed to predict resource consumption
of a computing task or method.  Narayanan~\emph{et
  al}.~\cite{Naray-wmcsa00} use historical application logging data to
predict the fidelity of an application, which decides its resource
consumption although they only consider selected aspects of device
hardware and application inputs.
Gurun~\emph{et al}.~\cite{Gurun-mobisys04} extend the Network Weather
Service (NWS) toolkit in grid computing to predict offloading but give
less consideration to local device and application profiles.

Early research work also extended programming language and runtime
middleware to run applications in distributed manner.  Adaptive
Offloading~\cite{adaptive} leverages Java's object oriented design to
partition a Java application with a modified JVM.
Coign~\cite{Hunt99-coign} converts an application built from COM
components into a distributable application.  R-OSGi~\cite{R-OSGi}
extends the centralized module management functionality supported by
the OSGi specification to enable an OSGi application to be
transparently distributed across multiple machines.  In contrast, we
avoid modification of the runtime, choosing to introduce simple Java
annotations to identify methods available for remote execution.

MAUI~\cite{maui} describes a system that enables energy-aware offload of mobile code to
infrastructure.  Their main aim is to optimize energy consumption of
the mobile device, by estimating and trading off the energy consumed
by local processing \emph{vs}. transmission of code and data for
remote execution.  Although they find that optimizing for energy
consumption often also leads to performance improvement, their
decision process considers only relatively coarse-grained information,
compared with the complex characteristics of the mobile environment.
MAUI is also similar to ThinkAir in that it provides method-level,
semi-automatic offloading of code. However, the programmer makes only
relatively coarse-grained decisions as to what should be offloaded,
while ThinkAir provides very fine-grained control while still making
the final offload decision based on profiled data to avoid
significantly degrading performance.

More recently, CloneCloud~\cite{clonecloud:2011} proposed
cloud-augmented execution using a cloned virtual machine (VM) image as
a powerful virtual device. Cloudlets~\cite{cloudlet1, cloudlet2} analyse use of a nearby resource-rich computer, or
cluster of computers, to which the smartphone connects over a wireless
LAN.  They argue against use of the cloud due to the higher latency
and lower bandwidth available when connecting.  In essence, they make
use of the smartphone simply as a thin-client to access local
resources, rather than using the smartphone's capabilities directly,
offloading only when required.  Paranoid Android~\cite{paranoid} uses
QEMU to run replica Android images in the cloud to enable multiple
exploit and attack detection techniques to run simultaneously with
minimal impact on phone performance and battery life.  The Virtual
Smartphone~\cite{wowmom} uses the Android x86 port to execute Android
images in the cloud efficiently on VMWare's ESXi virtualization
platform, although they do not provide any programmer support for
utilising this facility. ThinkAir shares the same design approach as previous works of using the smartphone VM image inside the cloud for handling computation offloading. Different from them, ThinkAir targets a commercial cloud scenario with multiple mobile users instead of computation offloading of a single user.  Hence, we focus not only on the offloading efficiency and convenience for developers, but also on the elasticity and scalability of the cloud side for the dynamic demands of multiple customers.

%% file: arch.tex
 \section{ThinkAir Architecture}
 \label{s:architecture}


\begin{figure}
\centering
\includegraphics[scale=.75]{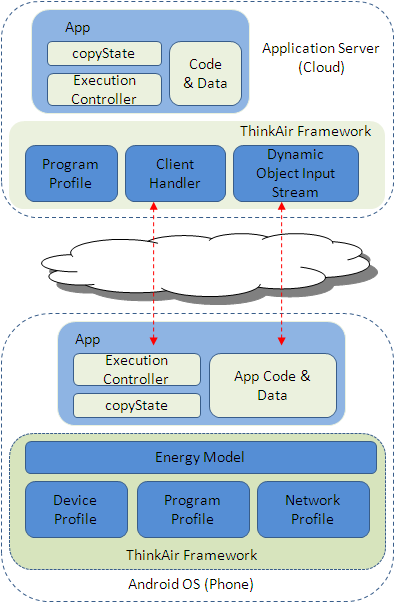}
\caption{\label{fig:framework}Overview of the ThinkAir framework.}
\end{figure}

The ThinkAir architecture is based on some basic assumptions which we
believe are already, or soon will become, true: \one~Mobile broadband
connectivity and speeds will continue to increase, enabling access to
cloud resources with relatively low Round Trip Times (RTTs) and high
bandwidths.  \two~As mobile device capabilities increase, so do the
demands placed upon them by developers, making the cloud an attractive
means to provide the necessary resources.  \three~The cloud will
continue to develop, supplying resources to users at low cost and
on-demand.

We reflect these assumptions in ThinkAir through four key concepts.

\one~\emph{Dynamic adaptation to changing environment}.
As one of the main characteristics of the mobile environment is rapid
change, the ThinkAir framework must adapt quickly and efficiently as
conditions change to achieve high performance as well as to avoid
interfering with the correct execution of the original software when
connectivity is lost.

\two~\emph{Ease of use for the developer}.
By providing a simple interface for developers, we both eliminate the
risk of misusing the framework and accidentally hurting performance
instead of improving it, and we allow less skilled and novice
developers to use it, increasing competition, one of the main driving
forces in today's mobile application market.

\three~\emph{Performance improvement through cloud computing}.
As the main focus of ThinkAir, we aim to improve both computational
performance and power efficiency of mobile devices by bridging
smartphones to the cloud.  If this bridge becomes ubiquitous, it will
serve as a stepping stone towards more sophisticated software.

\four~\emph{Dynamic scaling of computational power}.
To satisfy the customer's performance requirements for commercial grade service, we explore the 
possibility of dynamically scaling up and down the computational power 
at the server side. Like in Amazon EC2, the user has 
the possibility to choose the desired power of the server in our framework .
Furthermore, if the computation task can be parallelized, than the user can also ask
for more than one VM to execute his task in parallel.


The ThinkAir framework consists of three major components: the
execution environment~(\S\ref{s:execenv}), the application
server~(\S\ref{s:appserver}) and the profilers~(\S\ref{s:profilers}).
We will now give an overview of the framework, depicted in
Figure~\ref{fig:framework}, as a whole before describing each
component in detail.

The execution environment is accessed indirectly by the developer:
during development, they make only small modifications to class and
method definitions for those methods they believe may benefit from
offloading.  It is the compiler that introduces the code to interact
with the ThinkAir execution environment.  As the program runs, the
Execution Controller detects if a given method is a candidate for
offloading and handles all the associated profiling, decision making,
and communication with the application server \emph{without} the
developer needing to be aware of the details.

Currently implemented profilers consider device status (e.g.~WiFi and
cellular data connectivity, battery state, CPU load), program
parameters, execution time, network usage (i.e.~how much data would
have to be transmitted to make offloading a particular method
beneficial) as well as estimated energy consumption. The first time a
method is executed, only the environmental parameters, e.g.,~device
status and program parameters, are used to make the decision.  In
subsequent runs, other parameters are also used and their history
kept.

If the method is to be offloaded, it and its state are serialized and
sent to one or more cloud-hosted Application Servers for execution.
ThinkAir defines the protocol by which clients communicate with their
specific Client Handler, sending serialized method invocations and
receiving computed results.  The Client Handler receives execution
requests and the possible requests for additional computational power.
If there is no any special request for computational power, than it 
 inspects the requested method, loads any required
libraries (both native and Java), before executing the method itself
and returns any results or exceptions. Otherwise, if the client asks for
more resources than this clone owns, or asks for its task to be parallelized,
then the ClientHandler will resume the needed clones and collaborate 
with them on executing the task.
 Each application server is hosted in a virtualization environment in the cloud; for the
evaluation we report here, we used Oracle's VirtualBox virtualization
package,\footnote{\url{http://www.virtualbox.org/}} but any suitable
virtualization platform, e.g.,~Xen~\cite{xen} or QEMU~\cite{qemu}
would do.

%% file: execenv.tex
\section{Compilation and Execution}
\label{s:execenv}


In this section we will describe in detail the process by which a
developer writes code to make use of ThinkAir, covering the programmer
API and the compiler, followed by the execution flow including the
Execution Controller.  We will use a simple worked example throughout
to illustrate use of the framework.

\subsection{Programmer API}
\label{subsec:api}

ThinkAir provides a simple library that, coupled with the compiler
support, makes the programmer's job very straightforward.  Consider
the following code:

\begin{lstlisting}
public class CountingRandom {
  long count;

  public Long generate(long seed) {    	
    count++;
  	
    Random random = new Random(seed);
    return random.nextLong();
  }
}
\end{lstlisting}

This contains a single class \texttt{CountedRandom}, itself containing
a single method \texttt{generate} which the programmer wishes to
offload.  This method makes (somewhat trivial) use of a local counter
\texttt{count}.  As with any class and method to be offloaded, the
following steps must be performed:

\begin{itemize}
\item
The class is modified to extend the abstract class Remoteable, which
implements Serializable and is part of ThinkAir library.
\item
Methods which should be considered for offloading are annotated with
annotation ``$@$Remote''.
\item
The constructor creates a local \emph{ExecutionController} to
control the flow of program execution and act as a gate to the cloud
server.  One of these must be created per thread.
\end{itemize}

This provides enough information to enable the ThinkAir code generator
to be executed against the modified code.  This takes the source file
and generates the necessary remoteable method wrappers and utility
functions.  The modified code for our example is as follows:

\begin{lstlisting}
public class CountedRandom extends Remoteable {
  long count;

  public CountedRandom(ExecutionController ec) {
    this.controller = ec;
  }

  @Remote
  public Long generate(long seed) {   	
    count++;
  	
    Random random = new Random(seed);
    return random.nextLong();
  }
}
\end{lstlisting}

This modified code is then passed through our compiler,
\emph{Remoteable Code Generator}.  Following this, the final version of
the code, able to be offloaded, is as follows:

\begin{lstlisting}
public class CountedRandom extends Remoteable {
  long count;

  public CountedRandom(ExecutionController ec) {
    this.controller = ec;
  }

  public Long generate(long seed) {
    Method toExecute;
    Class<?>[] paramTypes = { long.class };
    Object[] paramValues = { seed };
    Long result = null;
    try {
      toExecute = this.getClass().getDeclaredMethod(
        "localGenerate", paramTypes);
      result = (Long) controller.execute(
        toExecute, paramValues, this);
    } catch (SecurityException e) {
      ...
    } catch (NoSuchMethodException e) {
      ...
    } catch (Throwable e) {
      ...
    }
    return result;
  }

  @Remote
  public Long localGenerate(long seed) {
    count++;

    Random random = new Random(seed);
    return random.nextLong();
  }

  @Override
  public void copyState(Remoteable state) {
  	CountedRandom localState = (CountedRandom) state;
  	this.count = localState.count;
  }
}
\end{lstlisting}

The \texttt{generate()} method is renamed to \texttt{localGenerate()}
and the original replaced by some Java reflection code whose job is to
invoke the method via the \emph{ExecutionController}, which can then
make the decision to offload or not, synchronizing state as necessary.
The \texttt{copyState()} method is generated to copy local state that
might have been changed during remote execution.  In this example the
value of local variable \texttt{count} is updated.



\subsection{Compiler}

A key part of the ThinkAir framework, the compiler comes in two parts:
the Remoteable Code Generator and the Customized Native Development
Kit (NDK).  The Remoteable Code Generator is a Java project that
translates the annotated code as described above.  Most current mobile
platforms provide support for execution of native code, for the
performance-critical parts of applications.  The Customized NDK exists
to provide native code support as cloud execution tends to be on x86
hosts while most smartphone devices are ARM-based.  To achieve this,
the Customized NDK simply uses the x86 support now unofficially
available in the distributed NDK to build all native libraries twice:
the first time for ARM as normal, the second time using a different
makefile to create x86 versions.  If this process fails for any
reason, then an instruction-level emulator could be deployed in the
application server environment; we do not consider this case further
here.



\subsection{Execution Controller}

The Execution Controller drives the execution of remoteable methods.
It decides whether to offload a method's execution, or to allow it to
continue locally on the phone. Its decision depends on data collected
about the current environment as well as that learnt from past
executions.

When a method is encountered for the first time, it is unknown to the
Execution Controller and so the decision is based only on
environmental parameters such as network quality.  If the connection
is of type WiFi, and the quality of connectivity is good, the
controller is likely to offload the method.  At the same time, the
profilers start collecting data.  If on a low quality connection, the
method is likely to be executed locally.

If and when the method is encountered subsequently, the decision on
where to execute it is based on the method's past invocations,
i.e.,~previous execution time and energy consumed in different
scenarios, as well as the current environmental parameters.
Additionally, the user also sets a policy according to their needs.
We currently define four such policies, combining execution time and
energy conservation:
\begin{itemize}
\item\textbf{None}.
The user chooses not to use the framework, causing all methods to be
executed locally.
\item\textbf{Execution time}.
Historical execution times are used in conjunction with environmental
parameters to prioritise fast execution when offloading,
i.e.~offloading only if execution time will improve (reduce) no matter
the impact on energy consumption.
\item\textbf{Energy}.
Past data on energy consumed energy is used in conjunction with
environmental parameters to prioritise energy conservation when
offloading, i.e.,~offloading only if energy consumption is expected to
improve (reduce) no matter the expected impact on performance.
\item\textbf{Execution time and energy}.
Combining the previous two choices, the framework tries to optimise
for both fast execution and energy conservation, i.e.,~offloading only
if both the execution time and energy consumption are expected to
improve.
\end{itemize}

Clearly more sophisticated policies could be expressed; discovering
policies that work well, meeting user desires and expectations is the
subject of future work.  Once the decision whether to offload or not
is taken, execution continues using Java reflection and the result is
sent back to the caller as detailed in the following section.

\subsection{Execution flow}
\label{subsec:exec-flow}

\begin{figure}[h!]
\centering
\includegraphics[scale=.6]{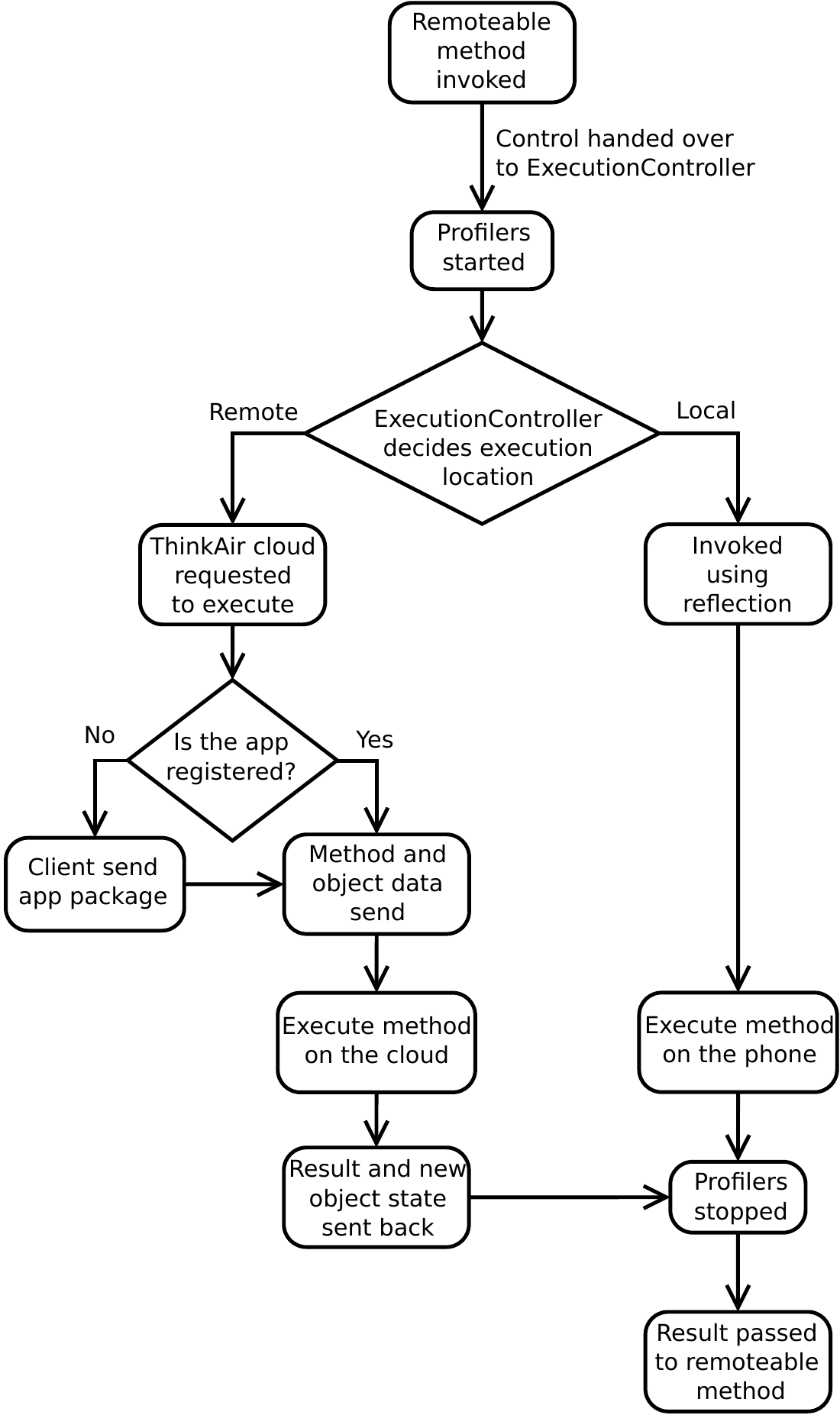}
\caption{\label{fig:remoteable-flowchart}Flow execution from
  calling a method to getting the result.}
\end{figure}

The result of the above compilation process is that, flow of control
is handed over to the Execution Controller when a remoteable method is
called as depicted in Figure~\ref{fig:remoteable-flowchart}.

On the phone, the Execution Controller first starts the profilers to
provide data for future invocations.  It then decides whether this
invocation of the method should be offloaded or not.  If it is, then
Java reflection is used to do so.  If not, then the calling object
must be sent to the application server in the cloud; the phone then
waits for results, and any mutated local state, to be returned.  If
the connection fails for any reason during remote execution, then the
framework falls back to local execution, discarding any data collected
by the profiler.  At the same time, the Execution Controller initiates
asynchronous reconnection to the server.  If an exception is thrown
during remote execution of the method then this is passed back in the
results and re-thrown on the phone, so as not to change the original
flow of control.


In the cloud, the Application Server manages clients that wish to
connect to the cloud, and this is covered in the following section.


%% file: appserver.tex
\section{Application Server}
\label{s:appserver}


The ThinkAir Application Server manages the cloud side of offloaded
code and is deliberately kept lightweight so that it can be easily
replicated.  It is started automatically when the remote Android OS is
booted, and consists of three main parts, described below: a client
handler, a dynamic object input stream, and the cloud infrastructure
itself.


\subsection{Client Handler}

The Client Handler executes the ThinkAir communication protocol,
managing connections from clients, and the process of receiving and
executing offloaded code, and returning results.


To manage client connections, the Client Handler registers when new
applications, i.e.,~new instances of the ThinkAir Execution
Controller, connect.  If the client application is unknown to the
application server, the Client Handler retrieves the application from
the client, and loads any class definitions and native libraries.  It
also responds to application-level \emph{ping} messages sent by the
Execution Controller as it measures connection latency. 

Note that an application may have more than one remoteable method; in
this way it is quite possible that a single Client Handler may end up
managing connections to more than one Execution Controller.  Each such
connection runs independently in a separate thread.  It is the client
(the phone) that remains responsible for ordering method invocations,
and any data sharing that results.  Extending this to enable
speculative execution of methods, introducing parallelization where
there previously was none, is a topic for future work.




Following the initial connection set up, the server waits to receive
execution requests from the client.  These consist of the necessary
data: the containing object, the requested method, the parameter types, 
the parameters themselves, and the possible request for extra computational power.
If there is no request for more computational power, then 
the Client Handler proceeds much
as the client would: the remoteable method is called using Java
reflection and the result, or exception if thrown, is sent back.
Well, there are some special cases regarding the exceptions.
As we will see later using a real application, if the exception is an \textit{OutOfMemoryError}
then the Client Handler will not send the exception to the client, 
but instead it will dynamically resume a more powerful clone, 
will delegate the task to him, get the result and send it back to the client.
If the user explicitly asks for more computational power, then again the Client Handler
will resume a more powerful clone to whom delegate the task.
In the same way, if the user asks for more clones
to execute his task in parallel, the Client Handler will resume the needed
clones, distribute the task among them, collect and give the results back 
to the client application.
Along with the return value, the Client Handler also sends some
profiling data to inform future offloading decisions made by the
Execution Controller.





\subsection{Dynamic Object Input Stream}

The \emph{ObjectInputStream} is part of the standard Java class
libraries available to Android.  It serves to deserialize Java objects
and primitive data types that have (typically) been saved using an
\emph{ObjectOutputStream}.  However, by default it simply throws an
exception (\emph{ClassNotFoundException} if an unknown class is
encountered.

Thus, to facilitate the creation of a completely open and generic
ThinkAir cloud, able to execute requests from any application created
for the framework, we introduce the \emph{DynamicObjectInputStream}.
This avoids the \emph{ClassNotFoundException} being thrown by being
able to request and load the Dalvik VM format Java bytecode
transmitted by the newly connected client.  In addition, it loads any
required native (x86) libraries retrieved from the client, these
having been generated by the Custom NDK at bulid time.




\subsection{Cloud Infrastructure}

To make the cloud infrastructure easily maintainable and to keep the
execution environment homogeneous in the face of, e.g.,~the
Android-specific Java bytecode format, we used a virtualization
environment allowing the system to be deployed where needed, whether
on a private or commercial cloud.  There are many suitable
virtualization platforms available,
e.g.,~Xen~\cite{xen}, QEMU~\cite{qemu} or Oracle's
VirtualBox.
In our evaluation we ran the Android x86
port\footnote{\url{http://android-x86.org/}} on VirtualBox.  To reduce
its memory and storage demand, we built a customized version of
Android x86, leaving out unnecessary components such as the user
interface or built-in standard applications.

\begin{table}
\centering\small
\begin{tabular}{lrrr}
\toprule
Type & CPUs & Memory (MB) & Heap Size (MB) \\
\midrule
basic & 1 & 200 & 32\\ [2pt]
main & 1 & 512 & 100\\ [2pt]
large & 1 & 1024 & 100\\ [2pt]
$\times$2 large & 2 & 1024 & 100\\ [2pt]
$\times$4 large & 4 & 1024 & 100\\ [2pt]
$\times$8 large & 8 & 1024 & 100\\ [2pt]

\bottomrule

\end{tabular}

\caption{\label{tab:vm-power}Different configurations of VMs.}
\end{table}

In our system, the users have 6 types of VMs with different configurations of CPU and memory to choose, which is shown in Table~\ref{tab:vm-power}. The VM manager can automatically scale up and down the computational power of the VMs and allocate more than one VMs for a task depend on the user requirement. The default setting for computation is only one VM with 1 CPU, 512MB memory, and 100MB heap size,  which clones the data and applications of the phone and we call it the \textit{primary server}. The main server is always online, waiting for the phone to connect to it. There is also a second type of VMs which can be of any configuration shown in Table~\ref{tab:vm-power}. This type of VMs in general does not clones the data and applications of a specific phone and can be allocated to any user on demand of compuational requirement and we call them the \textit{secondary servers}. The secondary servers  can be in any of these three states: \textit{powered-off}, \textit{paused}, or \textit{running}. When a VM is in powered-off state, it is not allocated any resources. The VMs in paused state is allocated the configured amount of memory, but they do not consume any CPU cycles. In the running state the VMs is allocated the configured amount of memory and will also make use of the CPU.

The Client Handler, which is in charge of the connection between the client (phone)
and the cloud, runs in the main server. The Client Handler is also in charge of the dynamic
control of the number of running secondary servers.
For example, if too many secondary VMs are running, it can decide 
to power-off or pause some of the VMs that are not executing any task.
Utilizing different states of the VMs has the benefit of controlling the 
allocated resources dynamically, but it also has the drawback of introducing the latency by resuming, starting, and synchronizing among the VMs.
From the experiments, we observed that the average time to resume one VM from the paused state is around 300ms. When the number of VMs to be resumed simultaneously is high (seven in our case), the resume time for some of the VMs can be upto 6 or 7 seconds because of the instant overhead introduced in the cloud. We are working on finding the best approach for removing this simultaneity  and stay in the limit of 1s for total resume time.
When a VM is in powered-off state, it takes on average 32s to start it, which is very high to use for methods that runs in the order of seconds.
However, there are tasks that takes hours to execute on the phone (for example Virus Scanning), for which it is still reasonable to spend
32s for starting the new VMs. An user may have different QoS requirements (e.g. complish time) for different tasks at different time, the VM manager needs to dynamically allocated the number of VMs to achieve the user expectation.

To make tests consistent, in our environment
all the virtual machines are run on the same physical server which is
a large multicore system with ample memory to avoid any effects of CPU
or memory congestion.
To simulate differences in connectivity between the local and remote
cloud we used three different mechanisms.  First with the VMs in the
same subnet as the WiFi connected phone, i.e.,~directly connected to
the access point; second, with the mobile client using an arbitrary
WiFi hotspot to connect to our local cloud over the Internet; and
finally, with the mobile client connecting over the Internet via the
3G data network.


%% file: profilers.tex
\section{Profiling}
\label{s:profilers}


The profilers are a critical part of the ThinkAir framework: the more
accurate and lightweight they are, the more correct offloading
decisions will be made, and the lower the overheads will be in making
them.  The profiler subsystem is highly modular so that it is
straightforward to add new profilers.  The current implementation of
ThinkAir includes three profilers (device, program, and network) which
feed into the energy estimation model, all of which we describe below.

For efficiency we use Android \emph{intents} to keep track of
important environmental parameters that do not depend on program
execution.  Specifically, we register listeners with the system to
track battery levels, and data connectivity presence, type (WiFi,
cellular) and subtype (GPRS, UMTS, \&c.).  This ensures that we do not
need to waste time or energy polling for the state of these factors.




\subsection{Device Profiler}

Since data from the Device Profiler will feed into the energy
estimation model we must consider how the application will behave when
using the ThinkAir framework.  In particular, CPU and the screen have
to be monitored whether or not a method is offloaded\footnote{We
  considered that simply turning off the screen during offloading
  would be too intrusive to users.}, but we must also monitor the WiFi
or 3G interfaces just when offloading.  These various components can
take the following states:


\begin{itemize}
\item[]\emph{CPU}.  The CPU can be \emph{idle} or have a utilization from
  1--100\% as well as two frequencies: 246\,MHz and 385\,MHz.
\item[]\emph{Screen}.  The LCD screen has a brightness level between
  0--255.
\item[]\emph{WiFi}.  The WiFi is either \emph{low} or \emph{high}.
\item[]\emph{3G}.  The 3G radio can be either \emph{Idle}, or in use
  with a \emph{Shared} or \emph{Dedicated} channel.
\end{itemize}

\subsection{Program Profiler}

The Program Profiler tracks a large number of parameters concerning
program execution.  After starting to execute a remoteable method,
whether locally or remotely, it uses the standard Android Debug API to
record:

\begin{itemize}
\item Overall execution time of the method.
\item Thread CPU time of the method, to discount the affect
  pre-emption by another process.
\item Number of instructions executed.\footnote{This required an
      adaptation of the distributed kernel due to what we believe is a
      bug in the OS using cascading profilers leading to inconsistent
      results and program crashes.}
\item Number of method calls.
\item Thread memory allocation size.
\item Garbage Collector invocation count, both for the current thread
  and globally.
\end{itemize}


\subsection{Network Profiler}

This is probably the most complex profiler as it must take into
account many different sets of parameters.  It combines both intent
and instrumentation-based profiling.  The former allows us to track
the network state so that we can e.g.,~easily initiate re-estimation
of some of the parameters such as RTT on network status change.  The
latter involves measuring the network RTT as well as the amount of
data ThinkAir sends/receives in a time interval, used to estimate the
\emph{perceived network bandwidth}.  This includes the overheads of
serialization during transmission, allowing more accurate offloading
decisions to be taken.


In addition, we track several other parameters for the WiFi and 3G
interfaces including number of packets transmitted and received per
second, uplink channel rate and uplink data rate for the WiFi
interface, and receive and transmit data rate for the 3G interface.
Doing so allows us to better estimate the current network performance
being achieved.

\subsection{Energy Estimation Model}

A key parameter for offloading policies in ThinkAir is the effect on
energy consumption.  
This requires dynamically estimating the energy consumed by methods
during execution.  We take inspiration from the recent PowerTutor~\cite{powertutor}
model which accounts for the CPU, LCD screen, GPS, WiFi, 3G and audio
interfaces on HTC Dream and HTC Magic phones.  The authors show that
the variation of estimated power on different types of phone is very
high, and present a detailed model for the HTC Dream phone which we
use in our experiments.  We have to modify their original model to accommodate
the fact that certain components, e.g.,~GPS and audio, have to be
operated locally and cannot be migrated to the cloud.



\begin{table}
\centering
\scalebox{0.85}{
\begin{tabular}{llll}
\toprule
\multirow{4}{*}{Model} & \multicolumn{3}{l}{$(\beta_{uh} \times freq_h + \beta_{ul} \times freq_l) \times util + \beta_{CPU} \times CPU_{on}$} \\ [2pt]
& \multicolumn{3}{l}{$+ \mbox{ } \beta_{Wifi_{l}} \times Wifi_l + \beta_{Wifi_h} \times Wifi_h$} \\ [2pt]
& \multicolumn{3}{l}{$ + \mbox{ }  \beta_{3G_{idle}} \times 3G_{idle} + \beta_{3G_{FACH}} \times 3G_{FACH}$} \\ [2pt]
& \multicolumn{3}{l}{$ + \mbox{ } \beta_{3G_{DCH}} \times 3G_{DCH}  + \beta_{br} \times brightness $} \\ [2pt]
\toprule
 Category & System variable & Range & Power coefficient\\
 \hline
\multirow{4}{*}{CPU} & \multirow{2}{*}{util} & \multirow{2}{*}{$1-100$} &$\beta_{uh}: 4.32$\\
  & & & $\beta_{ul}: 3.42$\\ [2pt]
  & freq$_l$, freq$_h$ & $0, 1$ & n.a.\\[2pt]
  & CPU$_{on}$ & $0, 1$ & $\beta_{CPU}: 121.46$\\[2pt]
  \hline
  \multirow{5}{*}{WiFi} & npackets, R$_{data}$ & $0 - \infty$ & n.a. \\[2pt]
  & R$_{channel}$& $1-54$ & $\beta_{cr}$ \\[2pt]
  & Wifi$_l$ & $0, 1$ & $\beta_{Wifi_{l}} : 20$ \\[2pt]
  & Wifi$_h$ & $0, 1$ &  $\beta_{Wifi_{h}} : ~approx 710$ \\[2pt]
 \hline
 \multirow{6}{*}{Cellular} & data\_rate & $0 - \infty$ & n.a. \\
   & downlink\_queue & $0 - \infty$ & n.a. \\[2pt]
   & uplink\_queue & $0 - \infty$ & n.a. \\[2pt]
  & 3G$_{idle}$ & $0, 1$ &  $\beta_{3G_{idle}} : 10$ \\[2pt]
  & 3G$_{FACH}$ & $0, 1$ &  $\beta_{3G_{FACH}} : 401$ \\[2pt]
  & 3G$_{DCH}$ & $0, 1$ &  $\beta_{3G_{DCH}} : 570$ \\ [2pt]
  \hline
  LCD & brightness & $0 - 255$ & $\beta_{br} : 2.40$ \\
\bottomrule
\end{tabular}
}
\caption{\label{tab:power-coeffs}Modified PowerTutor model for the HTC
  Dream Phone, dropping accounting for GPS and audio energy
  consumption.}
\end{table}

\begin{figure}
\centering
\includegraphics[scale=.5]{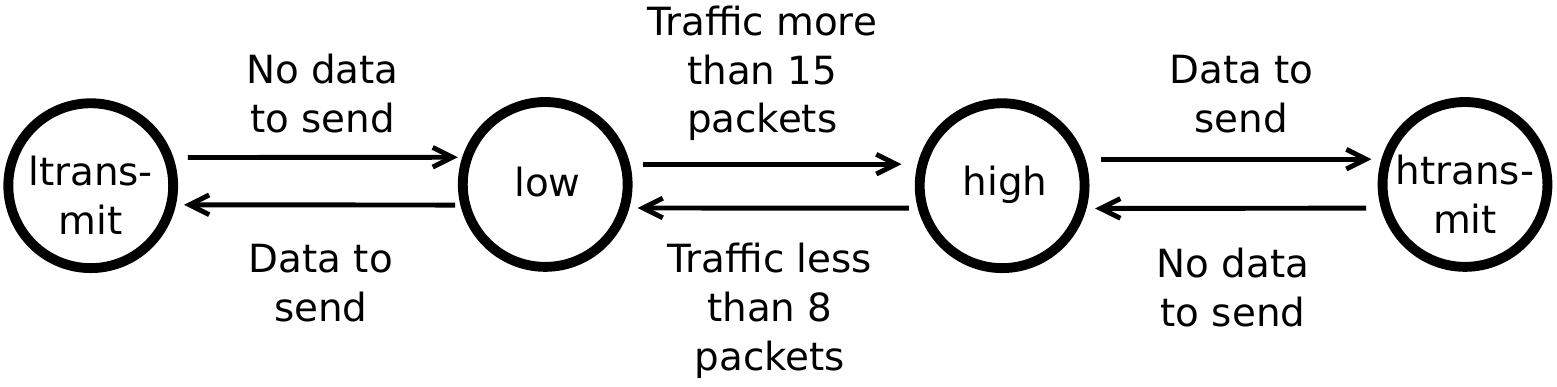}
\caption{\label{fig:wifi-interface}WiFi interface power states.}
\end{figure}

\begin{figure}
\centering
\includegraphics[scale=.48]{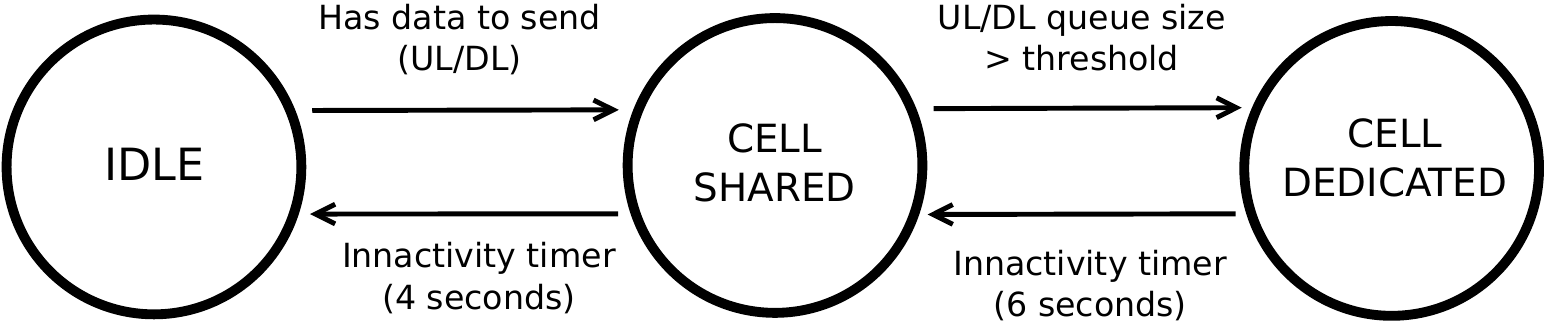}
\caption{\label{fig:radio-interface}3G interface power states.}
\end{figure}

By measuring the power consumption of the phone when it is at the
different cross products of the extreme power states,
e.g.,~considering just LCD and CPU, the different cross products are
[Full brightness, Low CPU] and [Low brightness, High CPU], the
PowerTutor authors found the maximum error to be $6.27\%$ if
individual components are assumed to be independent. This suggests
that a sum of independent component-specific power estimates is
sufficient to estimate system power consumption.  Thus, considering
each component in turn:
\\

\textbf{CPU}.  The key factors in CPU power consumption are CPU
utilization and frequency; the HTC Dream has two CPU frequencies,
246\,MHz and 385\,MHz, so we use the corresponding power coefficients
from the PowerTutor model, shown in Table~\ref{tab:power-coeffs}.
\\

\textbf{LCD}.  We use the PowerTutor values here, derived using a
training program to alter the screen's brightness from on to off.
\\

\textbf{WiFi}.  The WiFi power model is more complex than the others, taking into consideration the number
of packets transmitted and received per second ($n_{packets}$), and
the uplink channel and data rates ($R_{channel}$ and $R_{data}$
respectively).  The WiFi interface has four power states, depicted in
Figure~\ref{fig:wifi-interface}: low-power, high-power,
$l_{transmit}$, and $h_{transmit}$, entering the latter two only
briefly when transmitting data, returning to its previous power state
after sending data.  When transmitting at high data rates, the card is
only briefly in the transmit state (i.e.,~approximately 10--15\,ms per
second) and the time in the low-power transmit state is even shorter.
The WiFi component power consumption in either transmitting state is
approximately 1,000\,mW.  The low-power state is entered when the WiFi
interface is neither sending nor receiving data at a high rate and
power consumption in this state is 20\,mW.  In contrast, in the
high-power state the power consumption is approximately 710\,mW
depending on transmission parameters such as the number of packets
transmitted and received per second\footnote{Note that it is
  \emph{packet} rate not bit rate that determines the power
  state.}). Further details are presented in the original PowerTutor
paper~\cite{powertutor}.
\\

\textbf{Cellular}. The cellular interface power consumption model
depends on transmit and receive rates (data rates) and two queue
sizes, and distinguishes between the different cellular radio power
consumption modes using three key states of the communication channel
between base station and cellular interface~\cite{umts, 3g_radio}, as
depicted in Figure~\ref{fig:radio-interface}:

\emph{\idle}. In this state the cellular interface only receives
paging messages and does not transmit data.  Power consumption is
10\,mW.

\emph{\dedicated}. In this state, the cellular interface has a
dedicated channel for communication with the base station. It can
therefore use high-speed downlink/uplink packet access (HSDPA/HSUPA)
data rates, resulting in a power consumption of 570\,mW for the
cellular interface. When there is no activity for a fixed period of
time, the cellular interface enters the \shared{} state.

\emph{\shared}.  In this state the cellular interface shares
a communication channel to the base station. Its data rate is only a
few hundred bytes per second and therefore the cellular interface
power consumption in this state is 401\,mW. If there is a lot of data
to be transmitted, the cellular interface enters the \dedicated{}
state. Transition from \shared{} to \dedicated{} is triggered by
changes in the downlink/uplink queue sizes maintained for these two
states in the radio network controller. In the PowerTutor paper it is
indicated that state transition thresholds are 151\,bytes for the
uplink queue and 119 bytes for the downlink queue. Once either queue
size exceeds its threshold, \dedicated{} is entered. Otherwise, if
the interface is idle for a sufficient duration, the \idle{} state is
entered.
\\

We implement this energy estimation model inside the ThinkAir Energy
Profiler and use it to dynamically estimate the energy consumption of
each running method.  We present measurement results in the next
section.

%% file: eval.tex
\section{Evaluation}
\label{s:eval}

We evaluate ThinkAir using three sets of experiments. The first is
adapted from the Great Computer Language
Shootout.\footnote{\url{http://kano.net/javabench/}} They were
originally used to perform a simple comparison of Java
\emph{vs}. C++ performance, and therefore serve as a simple set of benchmarks
comparing local \emph{vs}. remote execution. The second is a more
recent set of benchmarks from the Computer Language Benchmark
Game~\cite{game}. Finally, we use five complete applications for a
more realistic evaluation: a sudoku solver, an instance of the
$N$-queens problem, a face detection program, a virus scanning, and
an image merging application.

\begin{figure}[bt!]
\centering
\includegraphics[scale=.7]{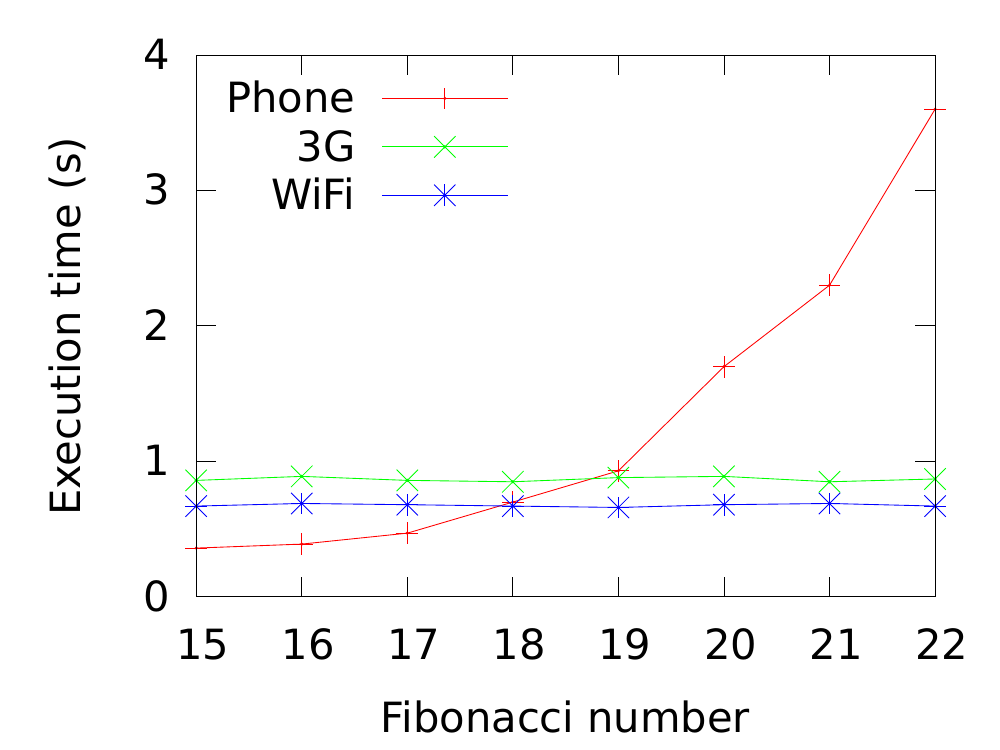}
\caption{\label{fig:biv}Boundary Input Value for \texttt{fibonacci(n)}}
\end{figure}

We define the \emph{boundary input value} (BIV) as the minimum value
of the input parameter for which offloading would give a benefit. We
use the \emph{Execution Time Policy} throughout so, for example, when
running \textit{Fibonacci(n)} under the execution time profile, we
find a boundary input value of $18$ when the phone connects to the
cloud through WiFi, i.e.,~execution of
\emph{Fibonacci(n)} is faster when offloaded for $n \geq 18$ (Figure \ref{fig:biv}).
The experiments are run under four different scenarios:

\begin{itemize}
\item \textbf{Phone}. Everything is executed on the phone.

\item \textbf{WiFi-Local}. The phone directly connects to the WiFi
router attached to the cloud server via the WiFi link.

\item \textbf{WiFi-Internet}. The phone connects to the cloud server
using a normal WiFi access point via the Internet.

\item \textbf{3G}. The phone is connected to the cloud using 3G.

\end{itemize}

Every result is obtained by running the program 20 times for every
scenario and averaged. Between two consecutive executions there is a
pause of 30\,seconds. The typical RTT of the 3G network that we used
for the experiments is around 100ms and that for the WiFi-local is
around 5\,ms. In order to test the performance of ThinkAir with
different quality of WiFi connection, we used both a very good
dedicated residential WiFi connection (RTT 50\,ms) and a commercial
WiFi hotspot shared by multiple users (RTT 200\,ms), which the users
may encounter on the move, for the WiFi-Internet setting. We did not
find any significant difference for these two cases, and hence we will
simplify them to a single case except for the full application
evaluations.

\subsection{Micro-benchmarks}
Originally used for a simple Java \emph{vs}. C++ comparison, each of
these benchmarks depends only on a single input parameter, making for
easier analysis. Results are shown in
Table~\ref{tab:complicated-benchmarks}. We find that, especially for
operations where little data needs to be transmitted, network latency
clearly affects the boundary value, hence the difference between
boundary values in the case of WiFi and 3G network connectivity. This
effect was also noted with Cloudlets~\cite{cloudlet1}. We also
include computational complexity of the core parts of the different
benchmarks, to show that with growing input values ThinkAir will only
become more efficient. Note that there are large constant factors
hidden by the $O$ notation, hence the different boundary input values
with the same complexity.




\begin{table}
\centering\small
\begin{tabular}{lrrlrr}
\toprule
Benchmark & \multicolumn{2}{c}{BIV} & Complexity & \multicolumn{2}{c}{Data (bytes)} \\
& WiFi & 3G & & Tx & Rx \\ \midrule
Fibonacci & 18 & 19 & $O(2^n)$ & 392 & 307 \\ [2pt]
Hash & 550 & 600 & $O(n^2log(n))$ & 383 & 293 \\ [2pt]
Hash2 & 3 & 3 & $O(nlog(n))$ & 361 & 300 \\ [2pt]
Matrix & 3 & 3 & $O(n)$ & 356 & 312 \\ [2pt]
Methcall & 2500 & 3100 & $O(n)$ & 338 & 297 \\ [2pt]
Nestedloop & 7 & 8 & $O(n^6)$ & 349 & 305 \\ [2pt]
Objinst & 2400 & 2700 & $O(n)$ & 337 & 296 \\ [2pt]
Sieve & 3 & 3 & $O(n)$ & 344 & 300 \\ [2pt]
\bottomrule
\end{tabular}
\caption{\label{tab:complicated-benchmarks}Boundary input values for
which it starts paying to offload, for WiFi and 3G connectivity, with
the computational complexity of the algorithms.}
\end{table}

\subsection{Realistic benchmarks}

The second set of benchmarks is similarly structured to the first one:
they depend on one input parameter and they have originally been used
for speed comparison of different programming languages. We perform
minimal modifications to make them work with ThinkAir. We describe
them as ``realistic'' as they range from binary tree operations to
regular expression matching to matrix calculations and simulation;
although not complete applications in their own right, these are the
types of operation that we feel might commonly be offloaded with
ThinkAir. Again, we present the boundary input values in
Table~\ref{tab:real-methods}.

\begin{table}
\centering\small
\begin{tabular}{lrrr}
\toprule
Benchmark & BIV & \multicolumn{2}{c}{Data (bytes)} \\
          &     & Tx & Rx \\
\midrule
binarytrees	 & 2 & 493 & 326\\ [2pt]
knucleotide  & 2 & 544 & 304 \\ [2pt]
mandelbrot   & 30 & 462 & 305 \\ [2pt]
nbody        & 310 & 929 & 896 \\ [2pt]
spectralnorm & 20 & 394 & 308 \\ [2pt]
\bottomrule

\end{tabular}

\caption{\label{tab:real-methods}Boundary input value of the real
methods for which it starts paying to offload using WiFi-Local. As
in Table~\ref{tab:complicated-benchmarks}, the results for 3G were
approximately the same.}
\end{table}

\subsection{Application benchmarks}

We consider five complete application benchmarks representative of
more complex and compute intensive applications: a Sudoku puzzle
solver, a solver for the classic $N$-Queens problem, a face
detection application, a Virus scanning application, and 
an application which combines two pictures into an unique large one.

\emph{Sudoku solver} Given a Sudoku configuration, try to solve it; return true if there is
a solution, and false otherwise.
\\
\begin{figure*}[!hbt]
\centering
\includegraphics[scale=0.55]{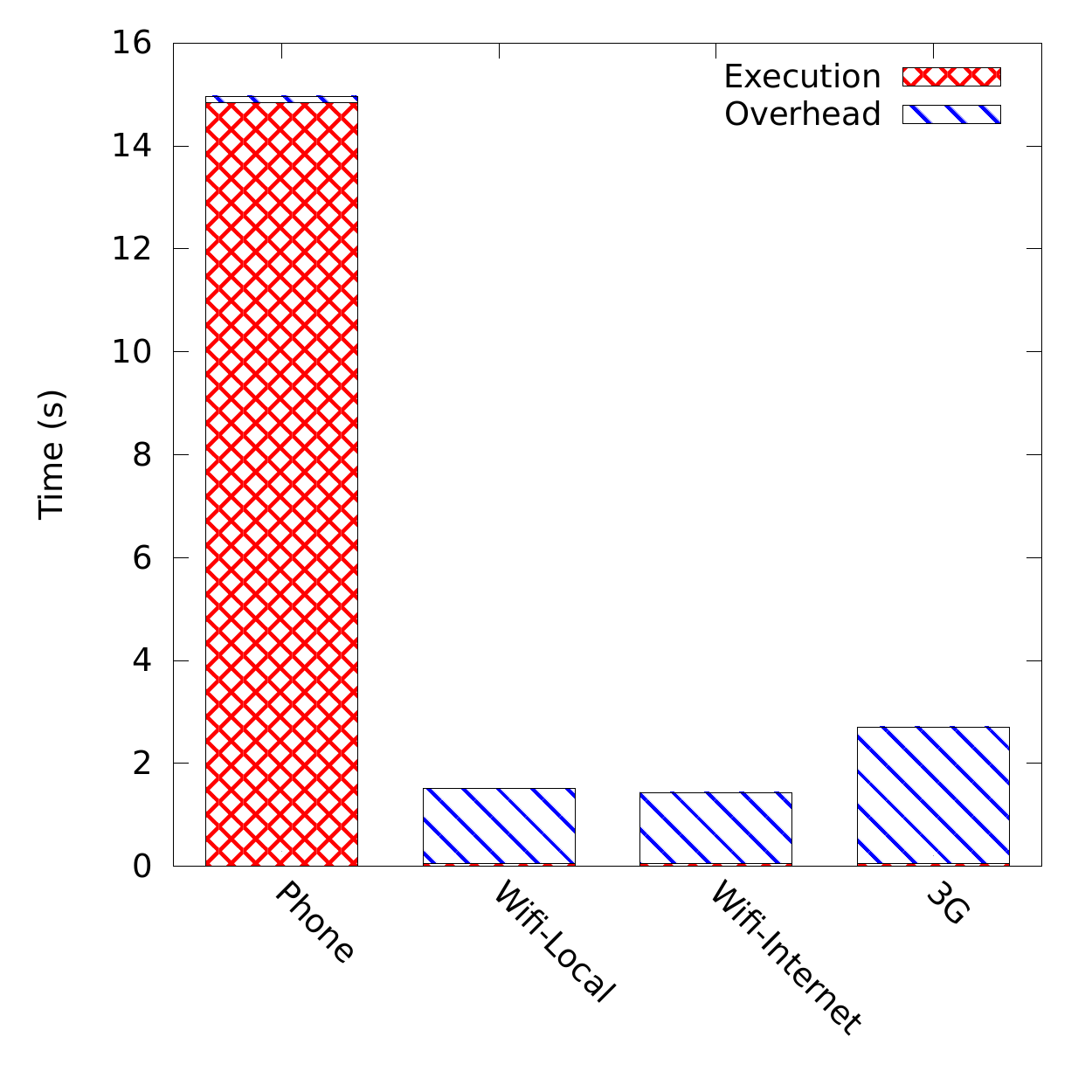}
\includegraphics[scale=0.55]{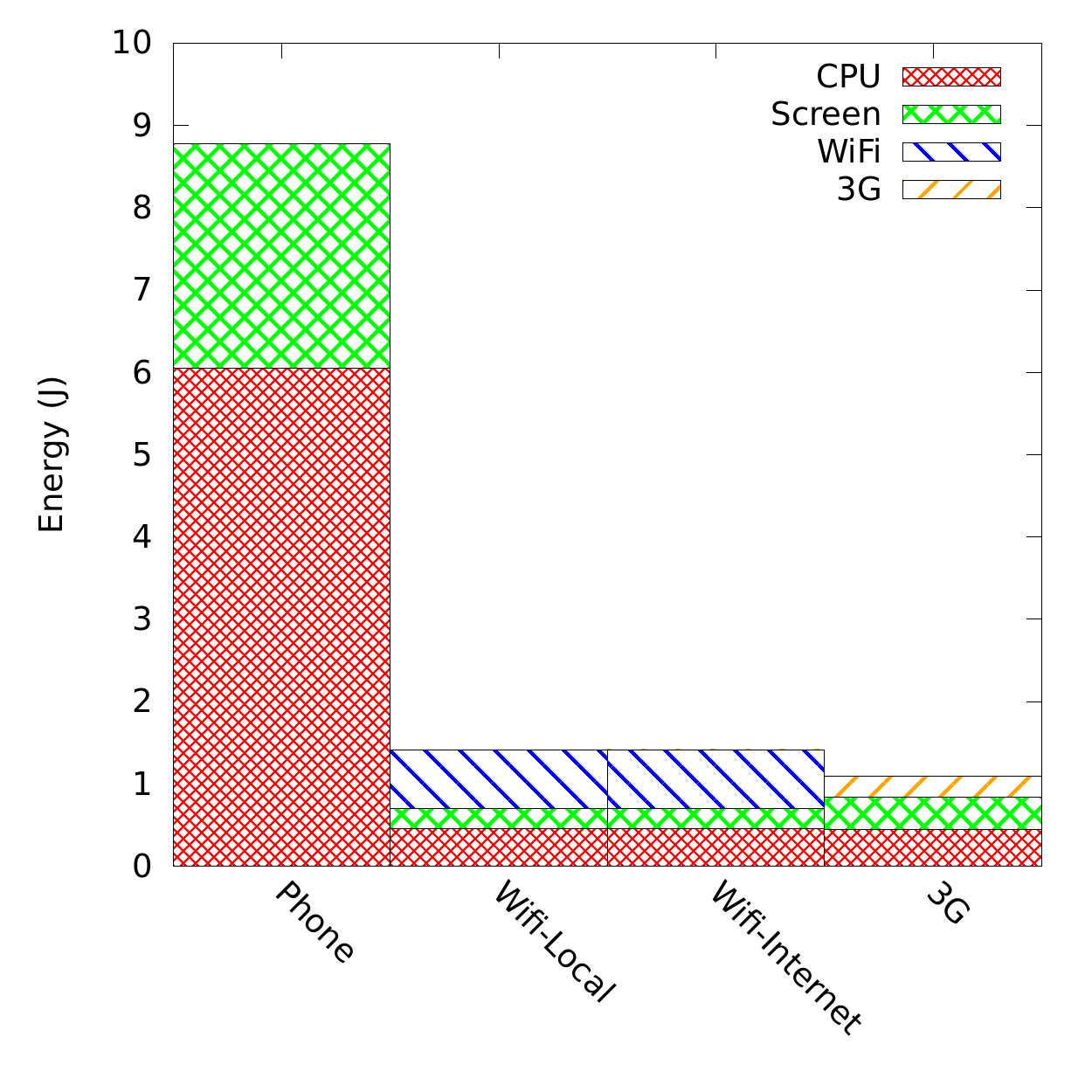}
\caption{\label{fig:sudoku-time-energy}Execution time and energy
consumption of the Sudoku solver.}
\end{figure*}
\begin{figure*}[!htbf]
\centering
\includegraphics[scale=0.55]{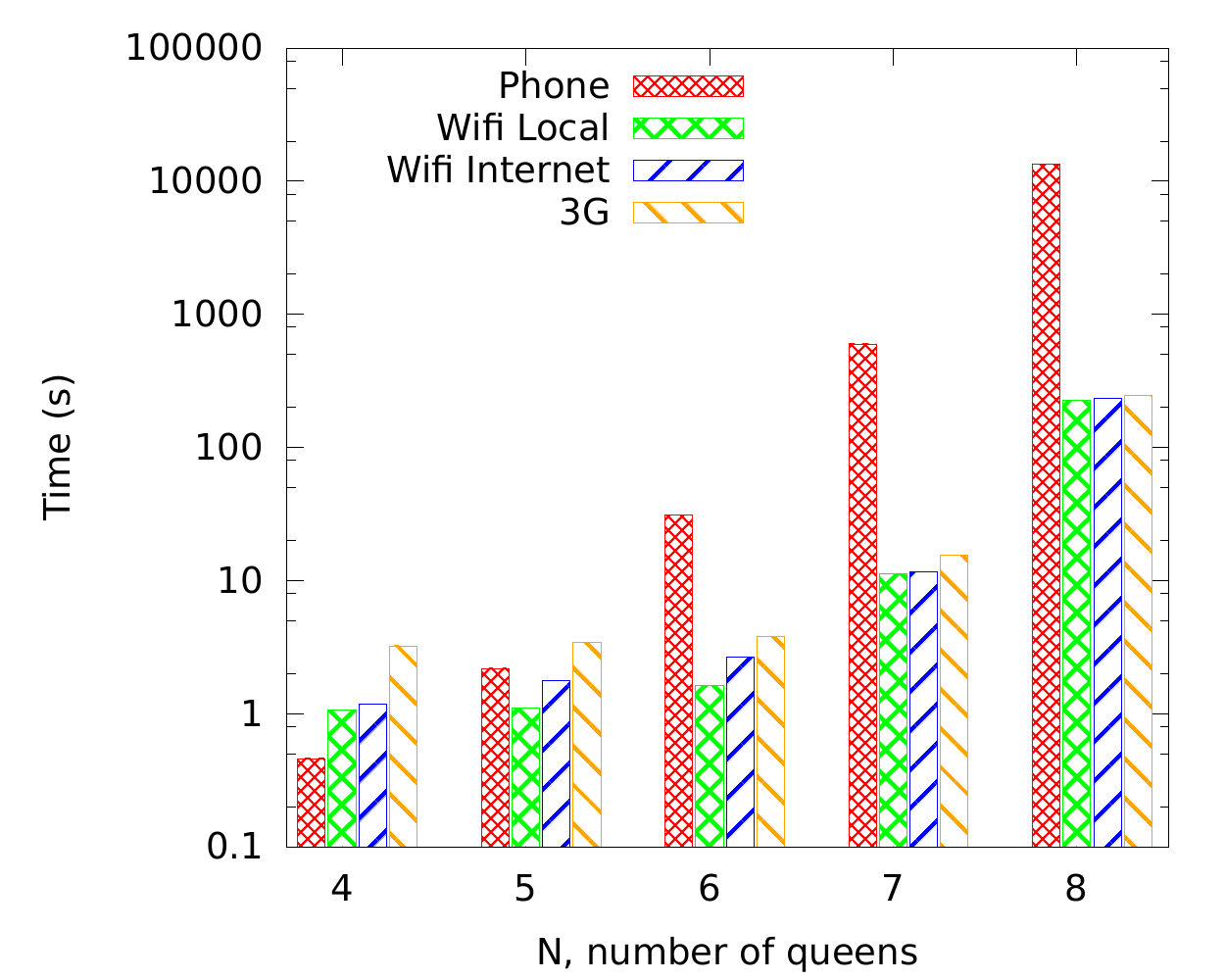}
\includegraphics[scale=0.55]{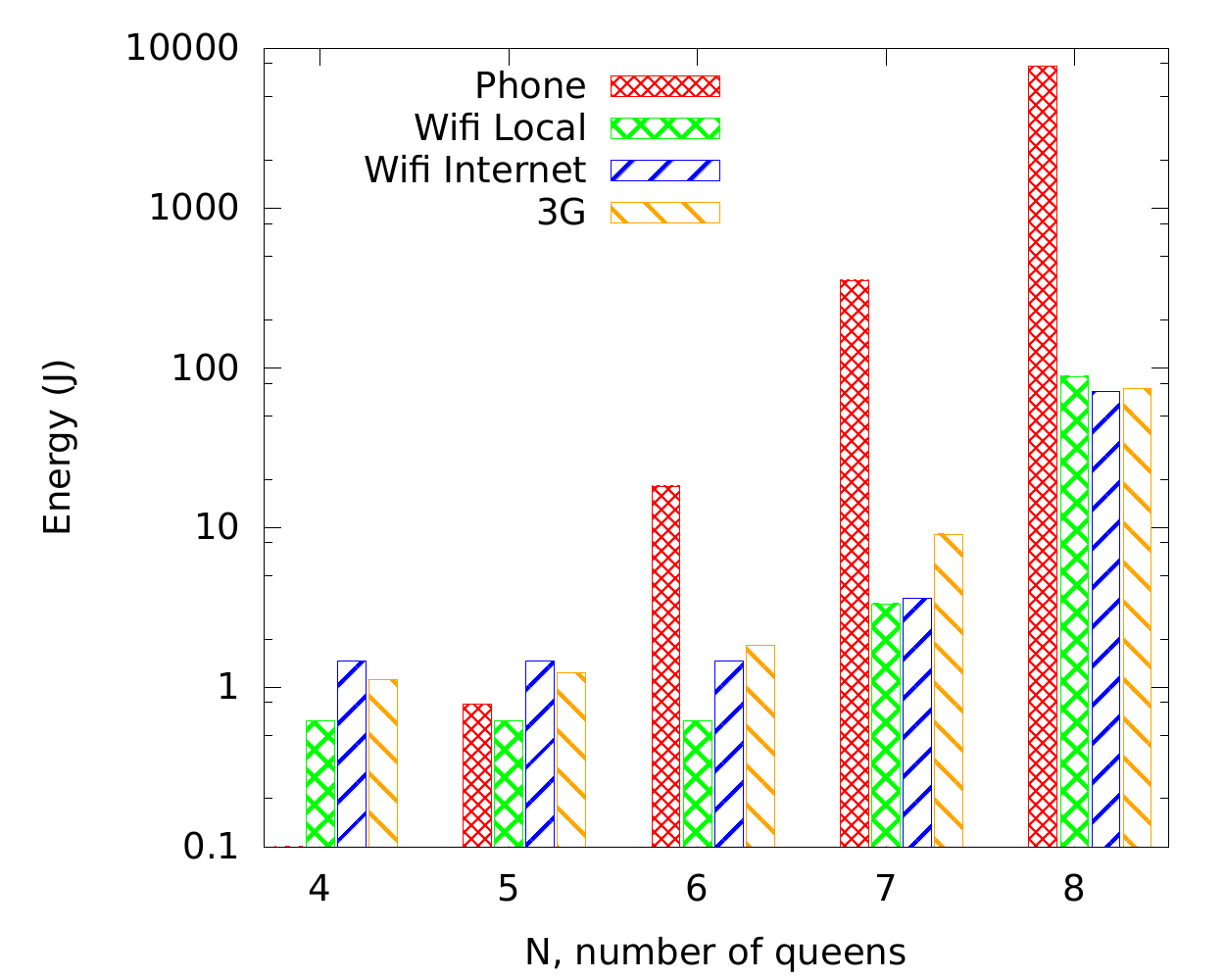}
\caption{\label{fig:n-queens-time-energy}Execution time and energy
consumption of the $N$-queens puzzle, $N = \lbrace 4, 5, 6, 7,
8\rbrace$.}
\end{figure*}
Figure~\ref{fig:sudoku-time-energy} shows the results for the Sudoku
Solver. We see that the execution time on the cloud is very much less
than on the phone, even though the overhead is substantially higher
due to the need to transmit and receive data. We can also see the
differences in the causes of energy consumption. When the method is
executed on the phone, energy consumption is very high due to both CPU
utilization (almost 100\% and always at the highest frequency) and the
fact that the screen remains on during execution. When offloading,
energy consumption is much lower: the extra energy consumed using the
radio interfaces to transmit and receive data is outweighed
by the reduction in energy consumed by the CPU and the screen.

\emph{$N$-Queens Puzzle}
An algorithm that finds all the solutions for the \emph{N}-Queens
Puzzle, returning the number of solutions found. We consider $4 \leq
\emph{N} \leq 8$ since at $N=8$ the problem becomes very
computationally expensive as there are 4,426,165,368 (i.e.,~64 choose
8) possible arrangements of eight queens on a $8 \times 8$ board, but only 92
solutions. We apply a simple heuristic constraining each queen to a
single column or row. Although this is still considered a brute force
approach, it reduces the number of possibilities to just $8^8 =
16,777,216$. We see from Figure~\ref{fig:n-queens-time-energy} that
for $N=8$ execution on the phone is
unrealistic as it takes hours to finish.
\begin{figure}
\centering
\includegraphics[scale=0.55]{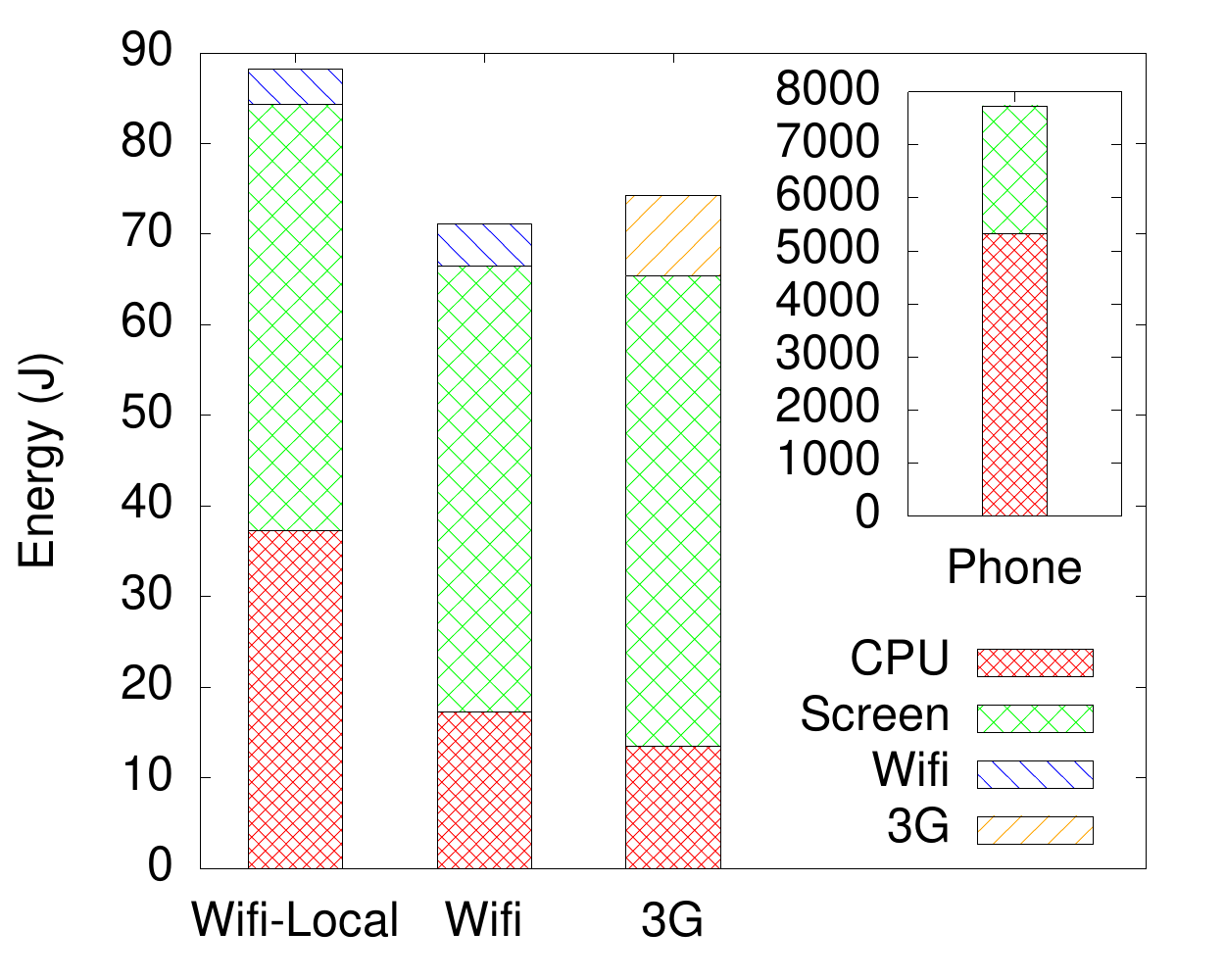}
\caption{\label{fig:8-queens-energy-details}Energy consumed by each
 component when solving 8-queens puzzle in different scenarios.}
\end{figure}
Figure~\ref{fig:n-queens-time-energy} again shows the time taken and
the energy consumed. We see that the boundary input value is between
5: for higher $N$, both the time taken and energy consumed in the
cloud are less than on the phone. In general, WiFi-Local is the most
efficient offload method although as $N$ increases, probably as higher
bandwidths lead to lower total network costs. Ultimately though,
computation costs come to dominate in all cases.

Figure~\ref{fig:8-queens-energy-details} breaks down the energy
consumption between components for $N=8$. As expected, when executing
locally on the phone, energy is consumed by the CPU and the screen, in
approximately the same proportion as with the Sudoku solver: again,
the CPU runs at approximately 100\% and at the highest possible
frequency throughout. When offloading, some energy is consumed by use
of the radio, and a slightly higher amount for 3G than WiFi. The
difference in CPU energy consumed between WiFi and WiFi-Local is due
to difference in the CPU speed of the local and cloud servers.

\emph{Face Detection}
Based on a third party
program,\footnote{\url{http://www.anddev.org/quick_and_easy_facedetector_demo-t3856.html}}
this is a simple face detection program that counts the number of
faces in a picture and computes simple metrics for each detected face
(e.g.,~distance between eyes). This demonstrates that it is
straightforward to apply the ThinkAir framework to existing code. The
actual detection of faces uses the Android API \emph{FaceDetector}, so
this is an Android optimized program and should be fast even on the
phone. We consider one run involving just a single photo and runs
involving comparing that photo against multiple (10, 100) others,
where the other photos have previously been loaded into the cloud
 e.g.,~comparing against photos from a user's Flickr
account. When running over multiple photos, we use the return values
of the detected faces to determine if the initial single photo is
duplicated within the set. In all cases, execution time and energy
consumed are much lower when executing on the cloud.

\begin{figure}
\centering
\includegraphics[scale=0.55]{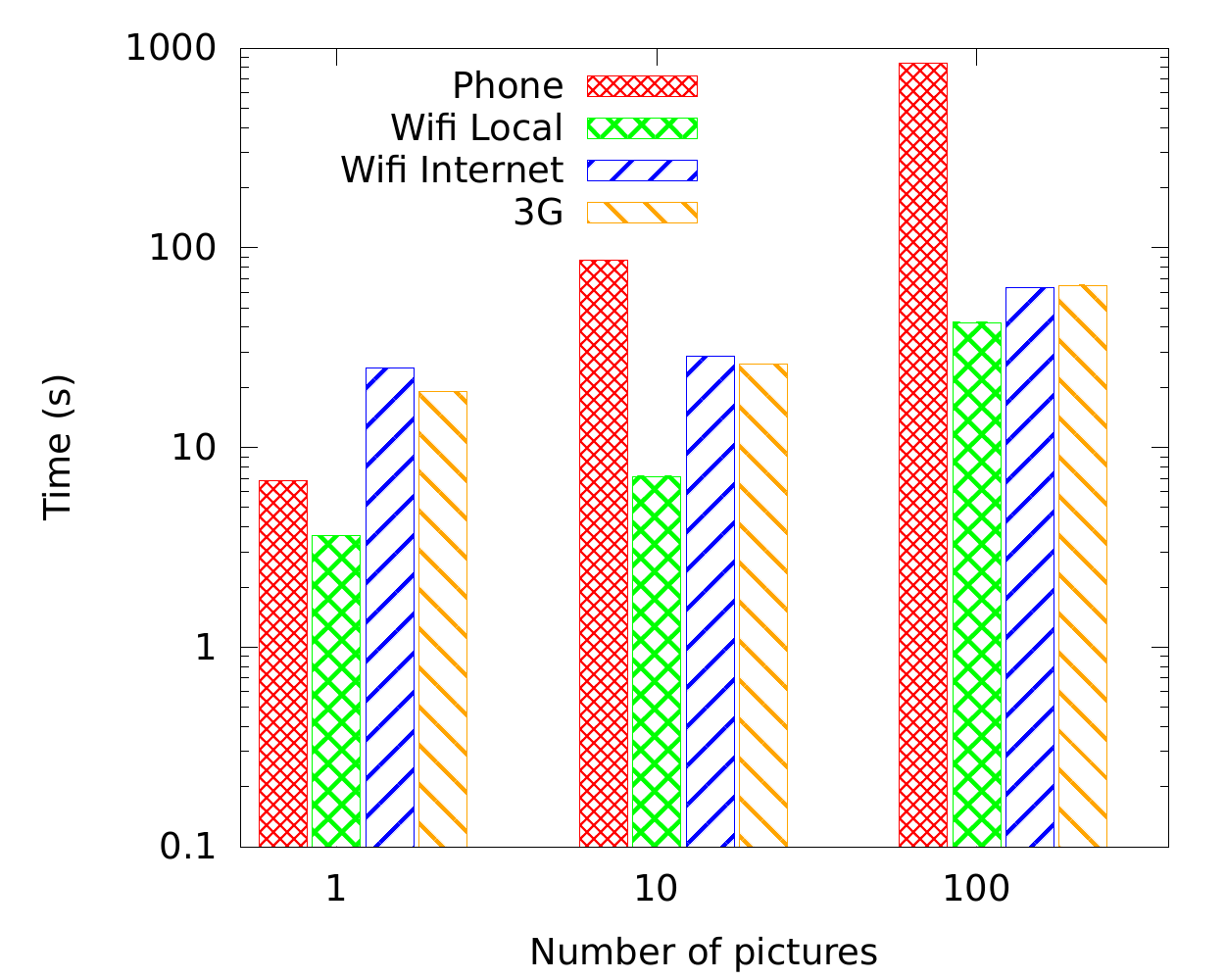}
\includegraphics[scale=0.55]{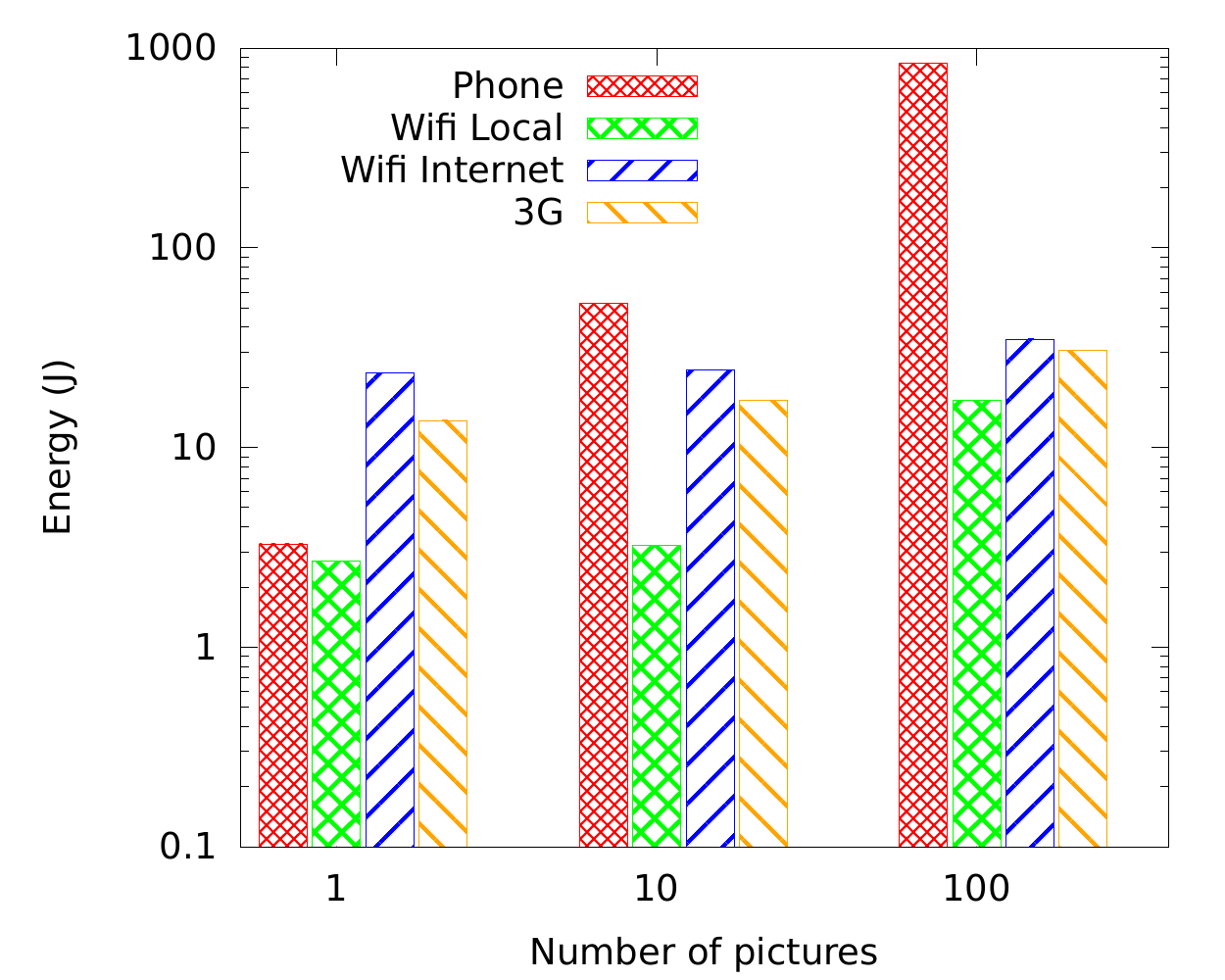}
\caption{\label{fig:face-time-energy}Execution time and energy
consumed for the face detection experiments.}
\end{figure}
\begin{figure}
\centering
\includegraphics[scale=0.55]{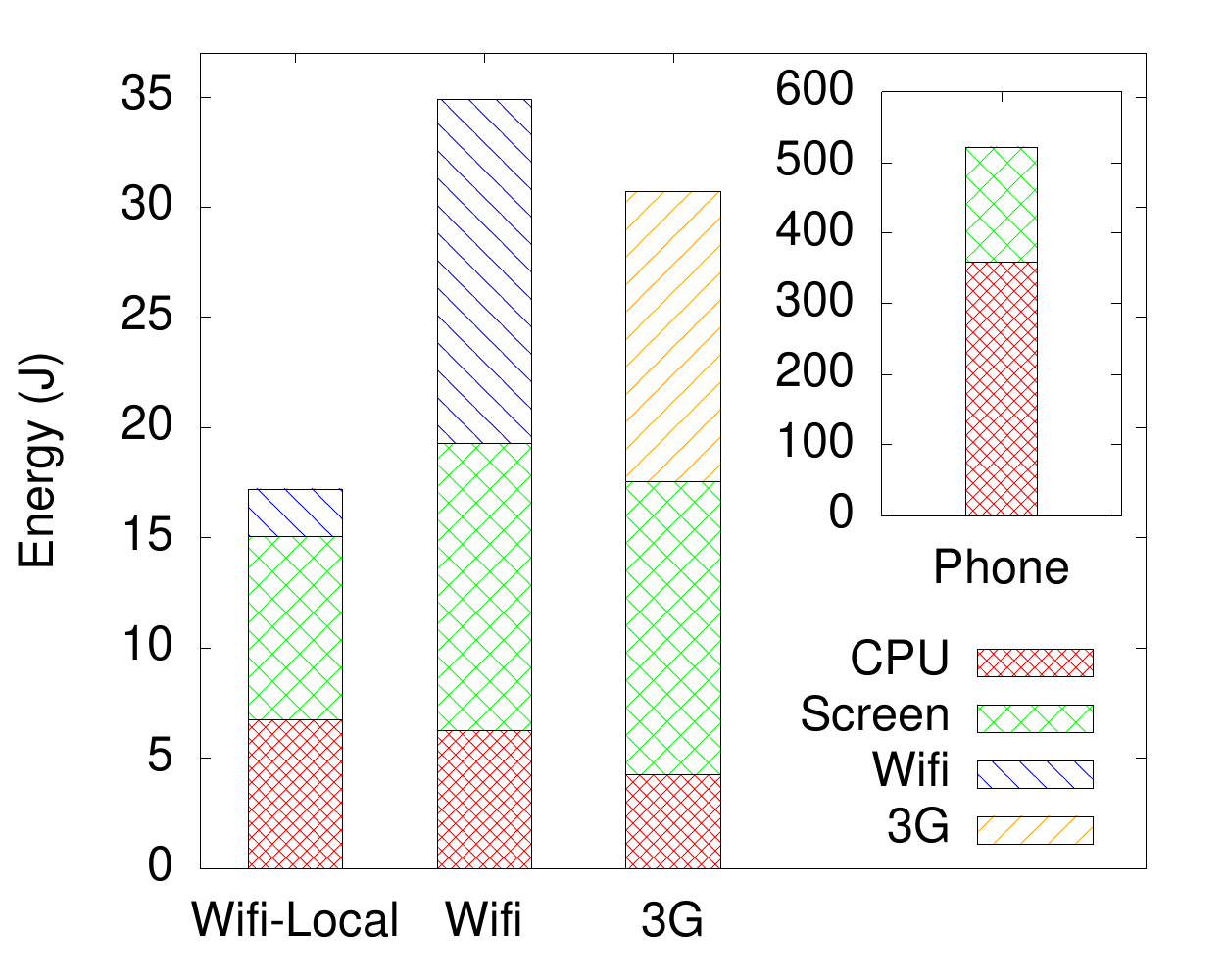}
\caption{\label{fig:100-face-energy-details}Energy consumed by each
component for face detection with 100 pictures in different scenarios.}
\end{figure}
Figure~\ref{fig:face-time-energy} shows the results for the face
detection experiments. The case where the face detection algorithm is
for just a single photo actually runs faster on the phone than
offloaded if the connectivity is not the best: as it is a
native API call on the phone and hence it is quite efficient.
However, as the
number of photos being processed increases, and in any case when the
connectivity is sufficiently high bandwidth and low latency, the cloud
proves more efficient once again.
Figure~\ref{fig:100-face-energy-details} shows the breakdown of the
energy consumed among components. As with the 8-Queens experiment
results shown in Figure~\ref{fig:8-queens-energy-details}, the
increased power of the cloud server compared with the local server
makes offloaded cases dramatically more efficient than the
case where everything is run locally on the phone.

\begin{figure*}[!tb]
\centering
\includegraphics[scale=0.55]{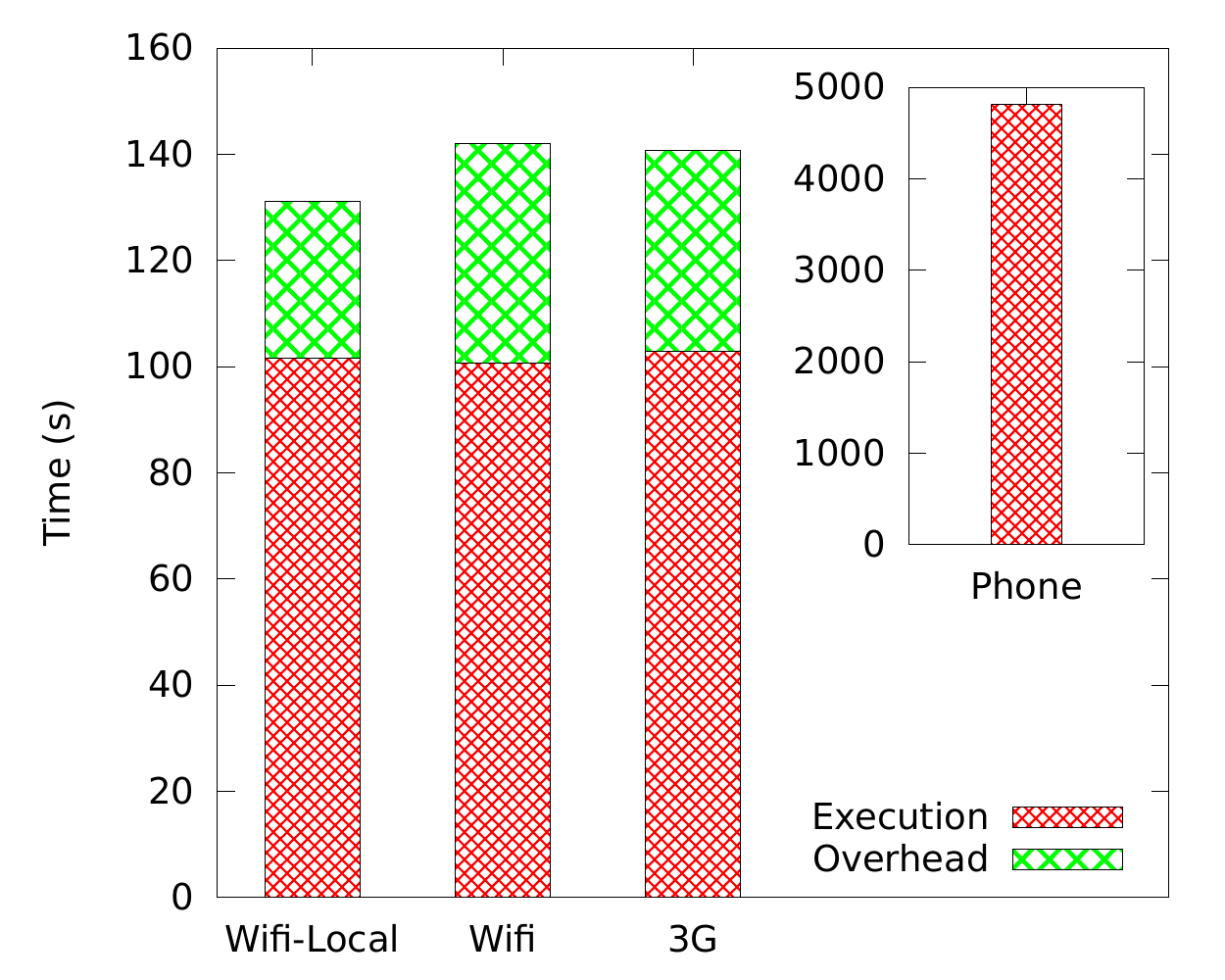}
\includegraphics[scale=0.55]{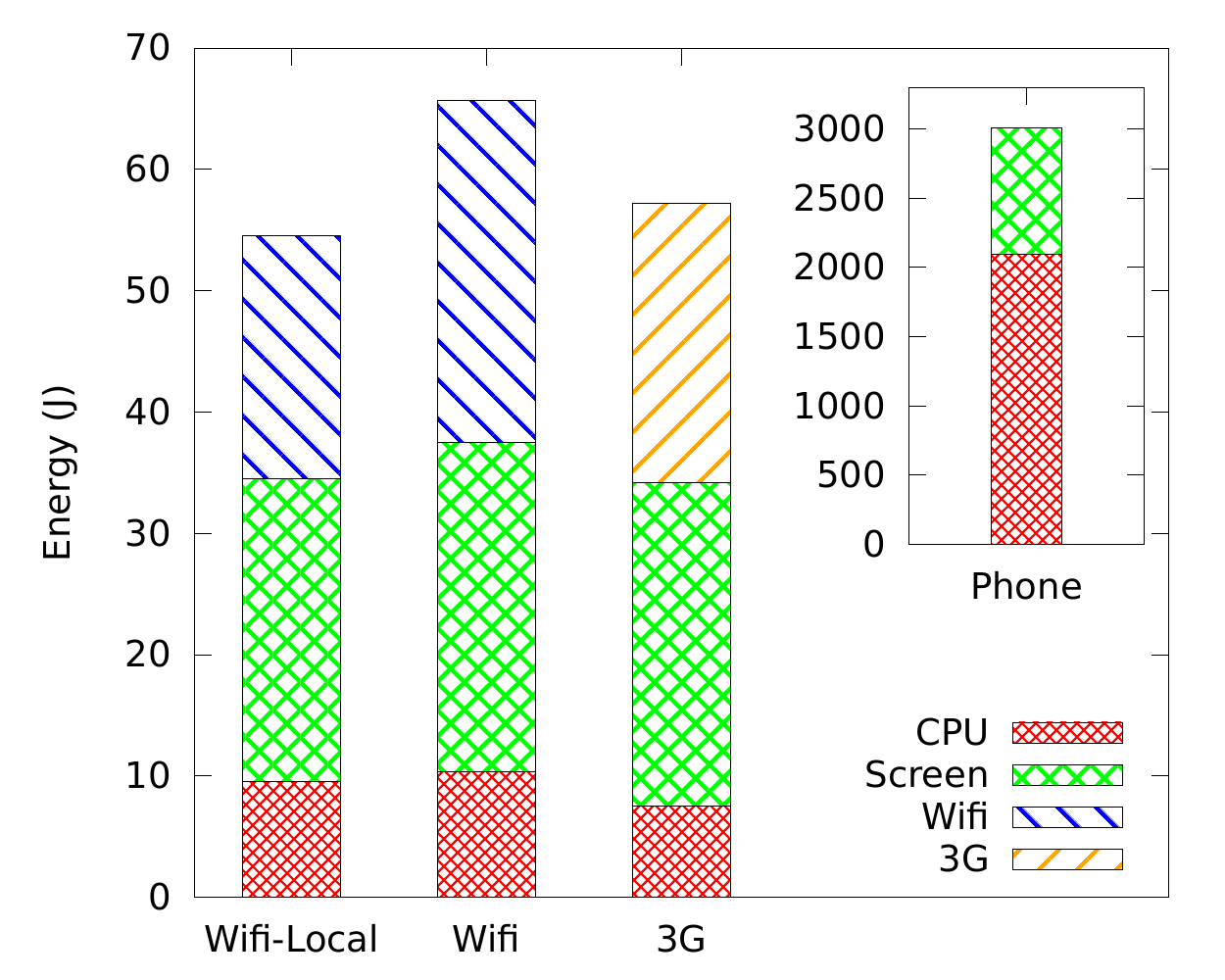}
\caption{\label{fig:virus-energy-time-details}Execution time and energy consumption of the virus scanning in different scenarios.}
\end{figure*}

\emph{Virus scanning}
We implement a virus detection mechanism for Android, which takes in a
database of 1000 virus signatures, the path to scan and returns the number of
viruses found. In our experiments, the total size of files in the directory is
10MB, and the number of files is around 3,500. We can see from
Figure~\ref{fig:virus-energy-time-details} that execution on the phone takes
more than one hour to finish, and it takes less than three minutes if offloaded.
In this figure we can also see the breakdown of the energy consumed by each
component. In this experiment the data to send for offloading is bigger compared
to the previous ones, so the comparison of the energy consumed by the WiFi and
3G is more fair. As a result we can say that WiFi is less energy efficient per
bit transmitted than 3G, which is also supported by the face detection
experiment (Figure~\ref{fig:100-face-energy-details}). Another interesting
observation is related to the energy consumed by the CPU. In fact, from the
results of all the experiments we can observe that the energy consumed by the
CPU is lower when offloading using 3G instead of WiFi.

\emph{Images combiner}
The intention of this application is to address the apps that cannot be run on
the phone due to lack of resources other than CPU. The Java VM heap size is a
big constraint for Android phones. If one application exceeds
16MB\footnote{\url{http://developer.android.com/reference/android/app/ActivityManager.html#getMemoryClass}}
of the allocated heap then it will throw an \textit{OutOfMemoryError}
exception\footnote{The maximum heap size can be configured from the phone
producers, so it can be different from the 16MB, which is the default on the
Android API}. Working with bitmaps in Android can be a problem if programmers do
not pay attention to memory usage. In fact, our application is a na\"{i}ve
implementation of combining two images next to each other into a bigger one. The
application takes in two images of size $(w_1, h_1)$, $(w_2, h_2)$ as input,
allocates memory for the final image of size $(\max\{w_1, w_2\}, \max\{h_1,
h_2\})$ and copies the content of each original image into the final one. The
problem here arises when the application tries to allocate memory for the final image,
resulting in \textit{OutOfMemoryError}, and making the execution impossible. We
are able to circumvent this problem by offloading the images to the cloud clone
and explicitly asking for high VM heap size. First, the clone will try to
execute the algorithm, but if does not have enough free VM heap size the
execution fails with \textit{OutOfMemoryError}. It will then resume a more
powerful clone and delegate the job to it. In the meantime, the application running on
the phone will free the memory occupied by the original images, and wait for the
final results. \begin{figure*}[!tb]
\centering
\includegraphics[scale=0.55]{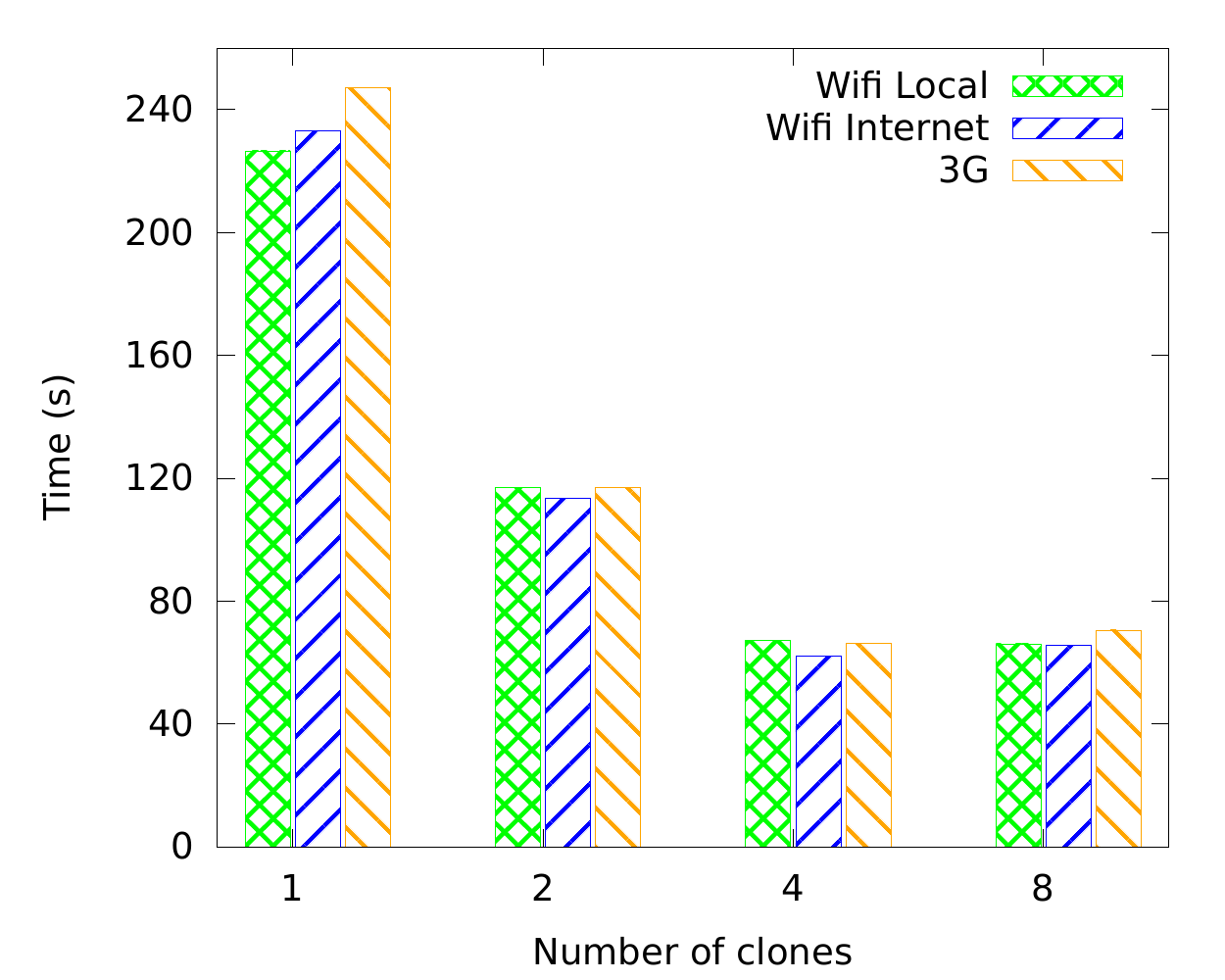}
\includegraphics[scale=0.55]{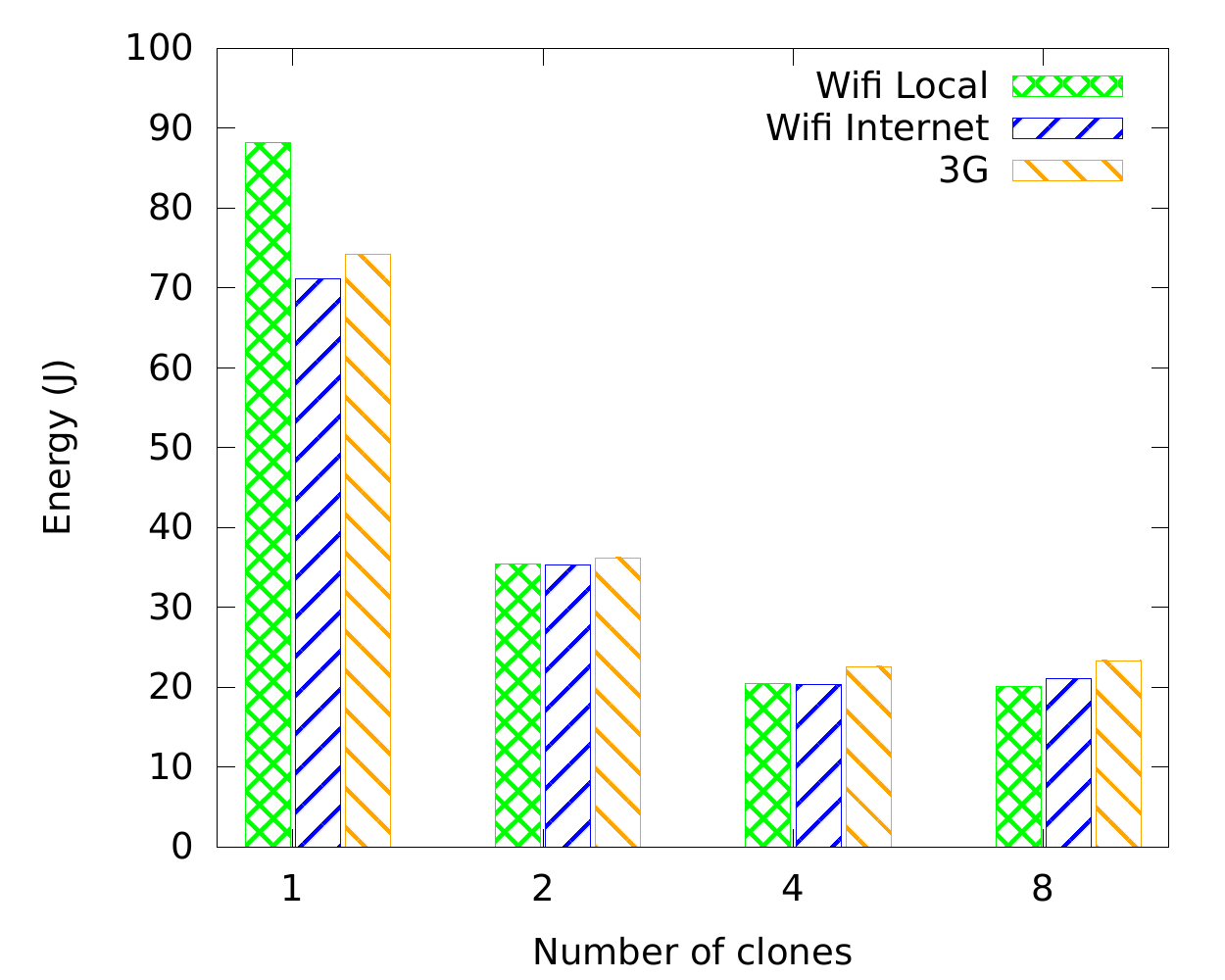}
\caption{\label{fig:8-queens-parallel-time-energy}Time taken and energy
consumed on the phone executing 8-queens puzzle using $N = \lbrace 1, 2, 4, 8\rbrace$ servers.}
\end{figure*}
\begin{figure*}[!tb]
\centering
\includegraphics[scale=0.55]{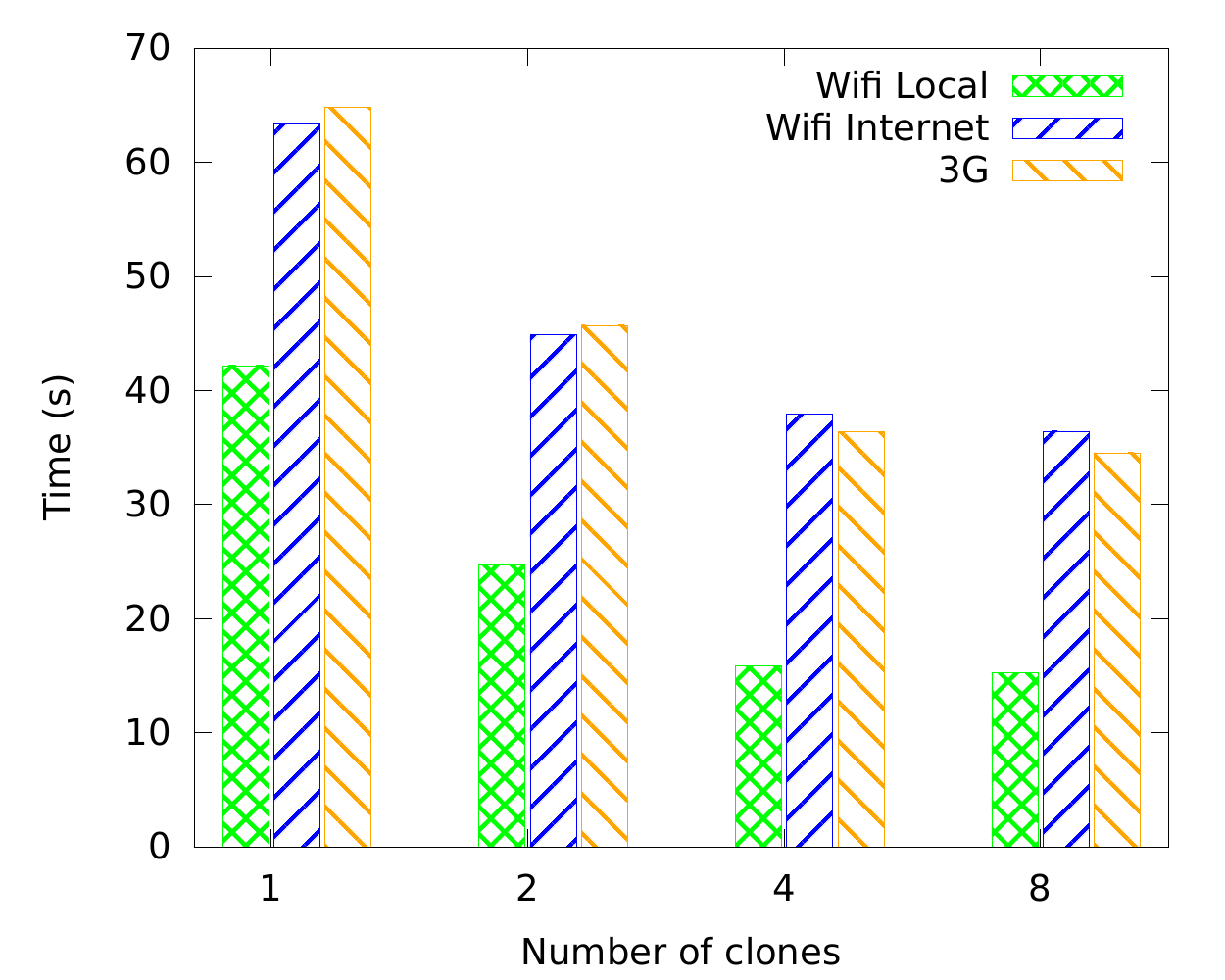}
\includegraphics[scale=0.55]{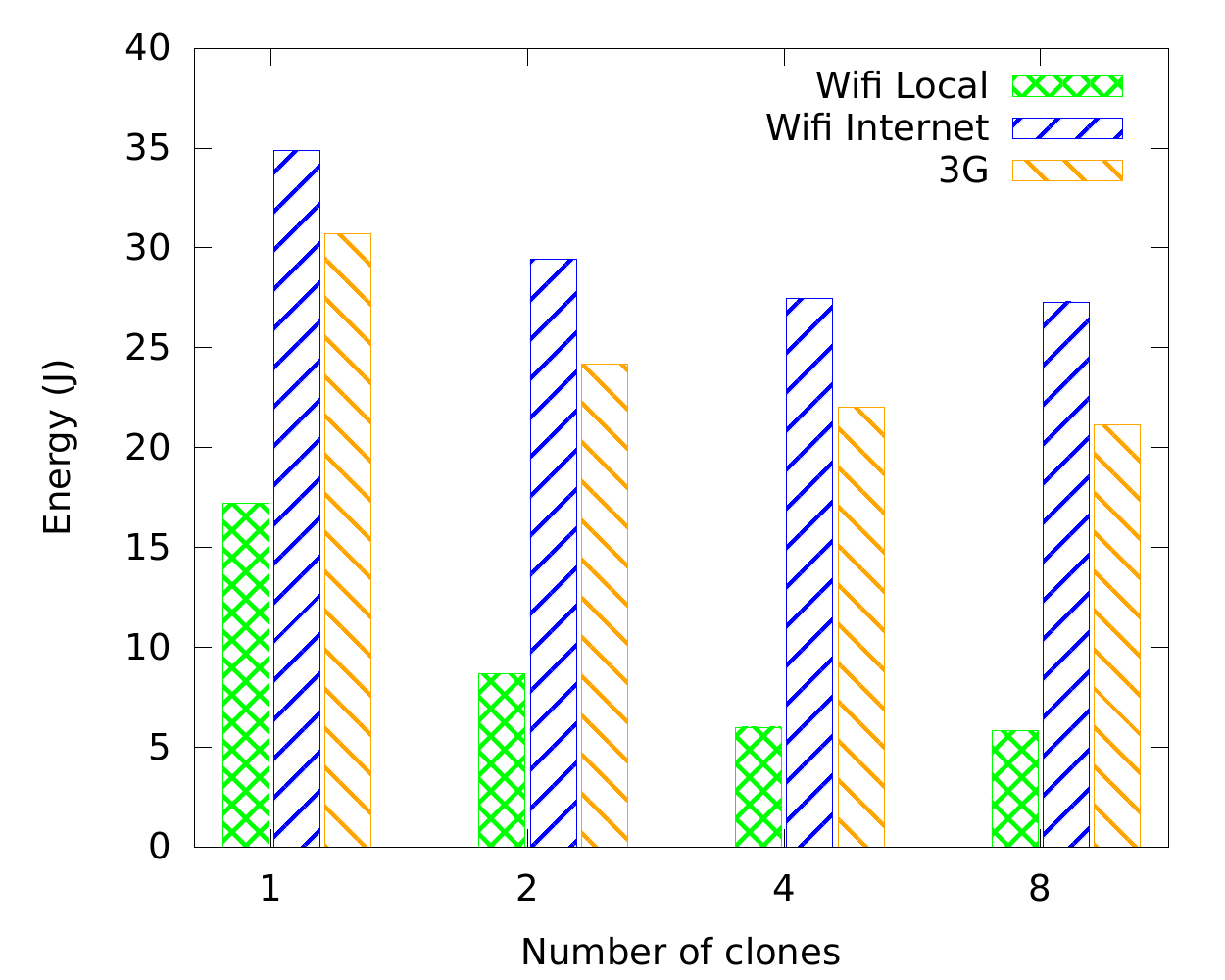}
\caption{\label{fig:100-faceDetection-parallel-energy-time}Time taken
and energy consumed for face detection on 100 pictures using
$N = \lbrace 1, 2, 4, 8\rbrace$ servers.}
\end{figure*}

\subsection{Parallelization with Multiple VM Clones}

In the last section, we showed that the framework can scale the processing power
up by resuming more powerful clones to delegate the task to. Another way of
achieving the scaling of the processing power is to exploit parallel execution.
If a user develops a parallelizable application, he can ask for more than one
clone to execute the task. In this section, we discuss the performance of three
complex applications, 8-Queens, Face Detection with 100 pictures, and Virus
Scanner using multiple cloud VM clones.
A single primary server communicates with the client and $k$ secondary clones,
$k \in \lbrace 1, 3, 7 \rbrace $. When the client connects to the cloud, it
communicates with the primary server which manages the secondaries, informing
them that a new client has connected. All interactions between the client and
the primary are as usual, but now the primary behaves as a (transparent) proxy
for the secondaries, incurring extra synchronization overheads. Usually the
secondary clones are kept in pause state to minimize the resources allocated.
Every time the client asks for service requiring more than one clone, the
primary server will resume the needed number of secondary clones. After the
secondaries finish their jobs, they are paused again by the primary server. The
time taken by a secondary clone to resume and connect to the main server is very
important, and it is included in the execution overhead.

The current modular architecture of the ThinkAir framework allows programmers to
implement any parallel algorithms with no modification to the ThinkAir code. In
our experiments, as the tasks are highly parallelizable, we evenly divide them
to be distributed to the secondaries.

In the 8-Queens puzzle case, the problem is split by allocating different
regions of the board to different clones and combining the results. For the face
detection problem, the 100 photos are simply distributed among the secondaries
for duplicates detection. In the same way, the files to be scanned for virus
signatures are distributed among the clones and each clone runs the virus
scanning algorithm on the files allocated. In all the following results, the
secondary clones are resumed from the paused state, and the resume time is
included in the overhead time, which in turn is included in the execution time.

\begin{figure*}[!tb]
\centering
\includegraphics[scale=0.55]{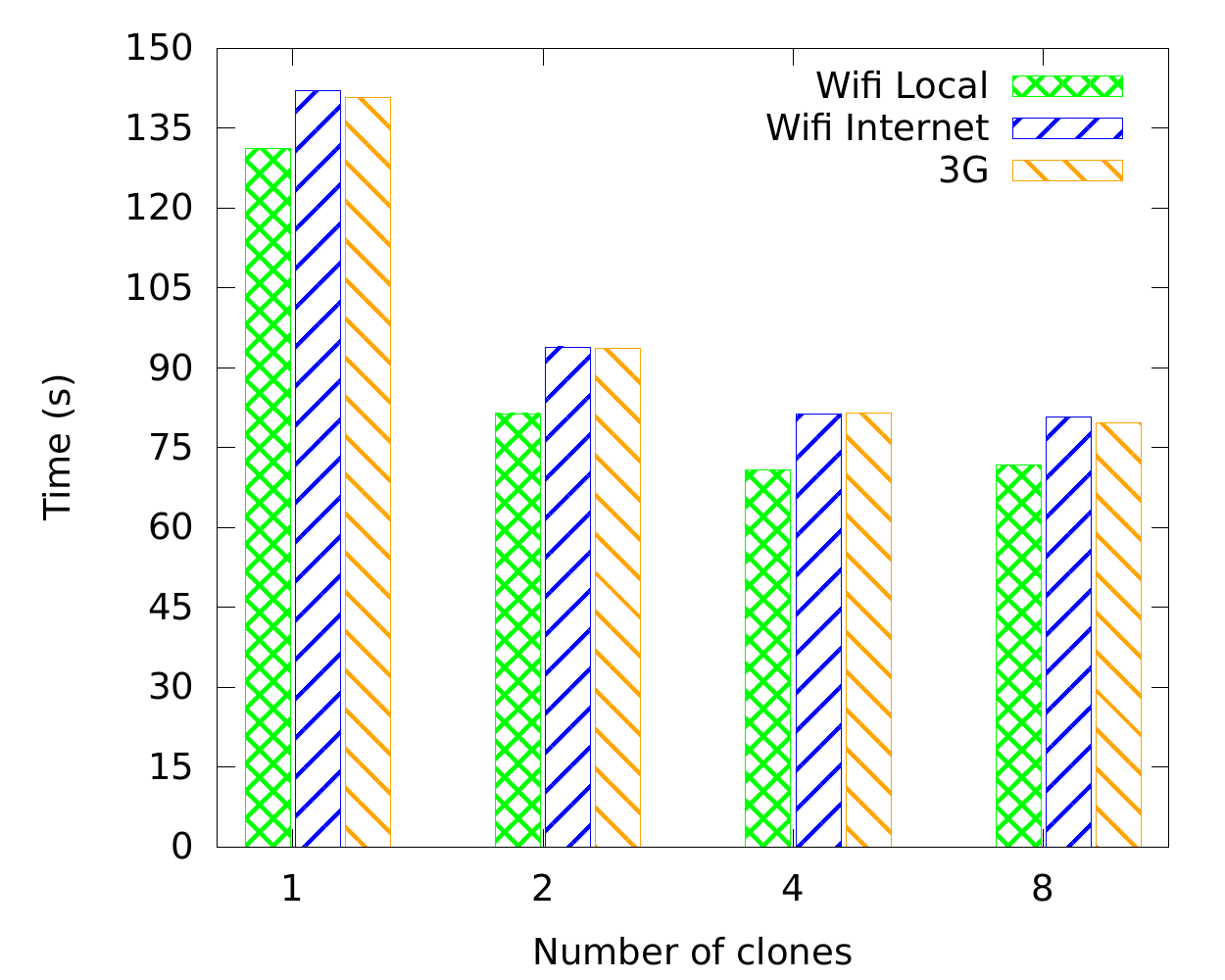}
\includegraphics[scale=0.55]{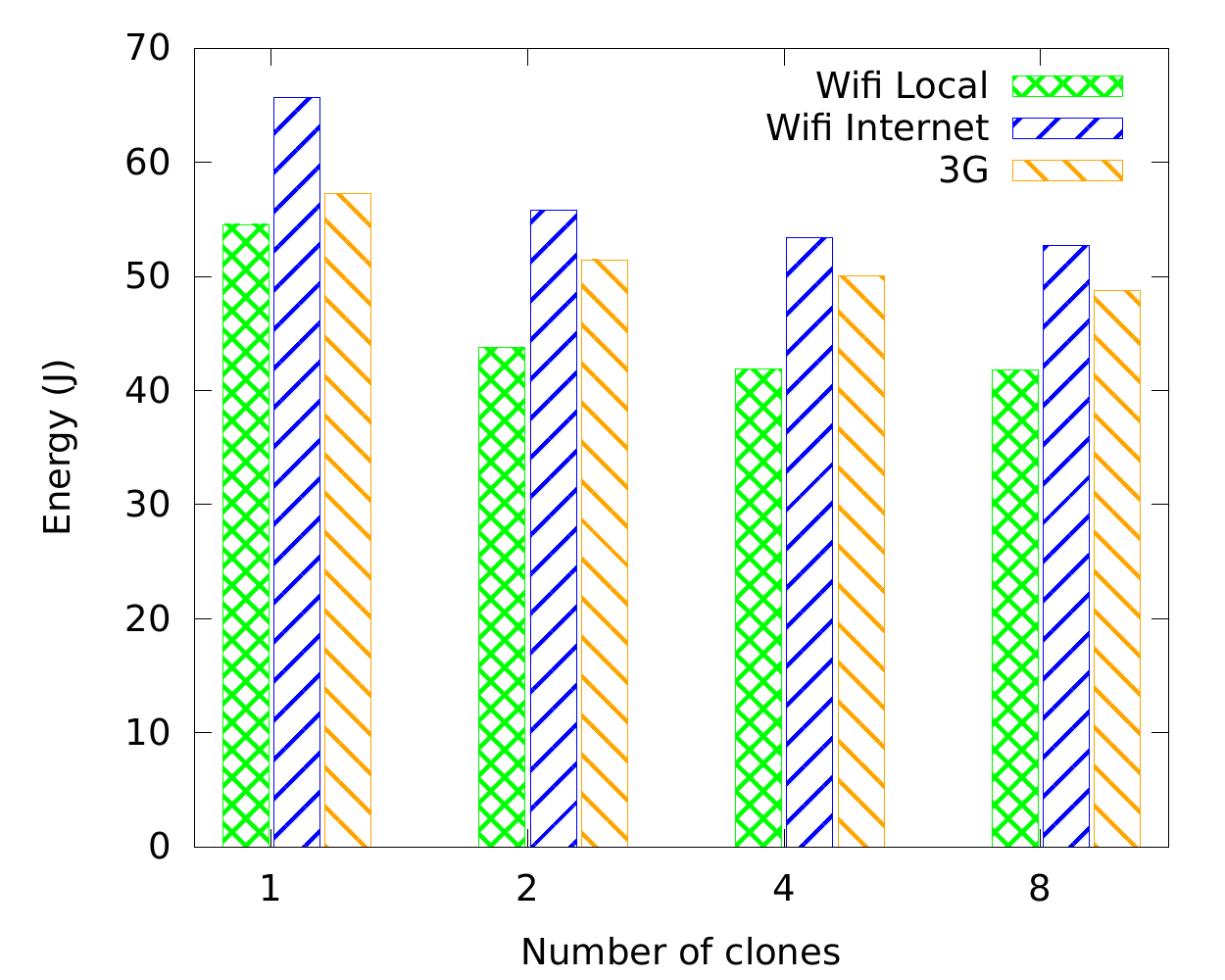}
\caption{\label{fig:virus-parallel-energy-time}Time taken
and energy consumed for virus scanning using $N = \lbrace 1, 2, 4, 8\rbrace$ servers.}
\end{figure*}

Figure~\ref{fig:8-queens-parallel-time-energy},
Figure~\ref{fig:100-faceDetection-parallel-energy-time}, and
Figure~\ref{fig:virus-parallel-energy-time} show the expected progression as the
number of clones increases. In the first case, almost all the benefit is
obtained with just 4 clones, since synchronization overheads start to outweigh
the running costs as the regions which the board has been divided to become very
small. The same effect is also observed in the other cases. Here one can also
see that the increased input size makes the WiFi less efficient in terms of
energy compared to 3G, which again supports our previous observations.

%% file: discuss.tex
\section{Discussion}
\label{s:discuss}

ThinkAir currently employs a conservative approach for data
transmissions: in addition to the method parameters and return values,
all data of the object encompassing the method is also transmitted.
This is obviously suboptimal as not all instance object fields are
accessed in every method and so do not generally need to be sent.  We
are currently working on improving the efficiency of data transfer for
remote code execution, combining static code analysis with data
caching.  The former eliminates the need to send and receive data that
is not accessed by the cloud.  The latter ensures that unchanged
values need not be sent, in either direction, repeatedly.  Note that
these optimization would need to be carefully applied however, as
storing the data between calls and checking for changes has large
overheads on its own.
               

ThinkAir assumes a trustworthy cloud server execution environment:
when a method is offloaded to the cloud, the code and state data are
not maliciously modified or stolen.  In our current ThinkAir
implementation, we also do not consider authentication of client
invocations of methods in the cloud.  We currently assume that the
remote server faithfully loads and executes any code received from
clients although we are currently working on integrating a lightweight
authentication mechanism into the application registration process.
Specifically, when the Client Handler in the cloud registers a new
application upon a request from an Execution Controller, it needs to
verify that the request is from a device that it can identify.  This
assumes pre-authentication between the client and the cloud.  For
example, a device agent can provide UI for the mobile user to register
the ThinkAir service before she can use the service. This registration
generates a shared secret based on user account or device identity,
which can be used to sign messages between the Execution Controller
and the Client Handler.

Privacy-sensitive applications may need more security requirements
than authentication. For example, if a method executed in cloud needs
private data from the device, e.g.,~location information or user
profile data, its confidentiality must be protected during
transmission.  For example, with encryption with a shared secret
between the Execution Controller and Client Handler.  We plan to
extend our compiler to support \texttt{SecureRemoteable} class to
support these security properties automatically and release the burden
from application developers.

%% file: concl.tex
\section{Conclusions}
\label{s:concl}

To conclude, we have presented ThinkAir, a framework for offloading
mobile computation to the cloud.  Using ThinkAir requires only simple
modifications to an application's source code by the programmer
coupled with use of the ThinkAir tool-chain.  Its evaluation
demonstrates the benefits of our approach to profiling and code
offloading, as well as accomodating changing computational requirements with the
ability of on-demand VM resource scaling and exploiting parallelism. We are
continuing development of several key
components of ThinkAir: we have ported Android to Xen allowing it to
be run on commercial cloud infrastructure, and we continue to work on
improving programmer support for parallelizable applications. Furthermore, we
see improving application parallelization support as a key direction to use the
capabilities of distirbuted computing of the cloud.

\section*{ACKNOWLEDGEMENTS}
\label{s:Ackn}

The authors would like to thank Jon Crowcroft, Steve Hand, Anil Madhavapeddy, Ranjan Pal, Julinda Stefa, Yu Xiao, and Collin Mulliner for their wonderful comments and insightful feedbacks.

%% file: thinkair.bbl
\begin{thebibliography}{10}

\bibitem{maui}
Eduardo Cuervo, Aruna Balasubramanian, Dae-ki Cho, Alec Wolman, Stefan Saroiu,
  Ranveer Chandra, and Paramvir Bahl.
\newblock {MAUI: making smartphones last longer with code offload}.
\newblock In {\em MobiSys '10: Proceedings of the 8th international conference
  on Mobile systems, applications, and services}. ACM, 2010.

\bibitem{clonecloud:2011}
Byung-Gon Chun, Sunghwan Ihm, Petros Maniatis, Mayur Naik, and Ashwin Patti.
\newblock Clonecloud: Elastic execution between mobile device and cloud.
\newblock In {\em Proceedings of the 6th European Conference on Computer
  Systems (EuroSys 2011)}, April 2011.

\bibitem{Aura}
J.P. Sousa and D.~Garlan.
\newblock Aura: an architectural framework for user mobility in ubiquitous
  computing environments.
\newblock In {\em Proc. of the 3rd Working IEEE/IFIP Conference on Software
  Architecture}, 2002.

\bibitem{Balan03-tactics}
Rajesh~Krishna Balan, Mahadev Satyanarayanan, SoYoung Park, and Tadashi Okoshi.
\newblock Tactics-based remote execution for mobile computing.
\newblock In {\em Proc. of The 1st International Conference on Mobile Systems,
  Applications, and Services}, pages 273--286, 2003.

\bibitem{Spectra}
R.~Balan, J.~Flinn, M.~Satyanarayanan, S.~Sinnamohideen, and H.~Yang.
\newblock The case for cyber foraging.
\newblock In {\em Proc. of the 10th ACM SIGOPS European Workshop}, 2002.

\bibitem{Odyssey}
B.~Noble, M.~Satyanarayanan, D.~Narayanan, J.~Tilton, J.~Flinn, and K.~Walker.
\newblock Agile application-aware adaptation for mobility.
\newblock In {\em Proc. of the ACM Symposium on Operating System Principles
  (SOSP)}, 1997.

\bibitem{Porras-foraging}
O.~Riva J.~Porras and M.~D. Kristensen.
\newblock {\em Dynamic Resource Management and Cyber Foraging}, chapter
  Middleware for Network Eccentric and Mobile Applications.
\newblock Springer Press, 2008.

\bibitem{dimorphic}
Andres Lagar-Cavilla Niraj, Niraj Tolia, Rajesh Balan, Eyal~De Lara, and
  M.~Satyanarayanan.
\newblock Dimorphic computing.
\newblock Technical report, 2006.

\bibitem{Berkeley-cloud-view}
Michael Armbrust, Armando Fox, Rean Griffith, Anthony~D. Joseph, Randy~H. Katz,
  Andrew Konwinski, Gunho Lee, David~A. Patterson, Ariel Rabkin, Ion Stoica,
  and Matei Zaharia.
\newblock Above the clouds: A berkeley view of cloud computing.
\newblock Technical Report UCB/EECS-2009-28, EECS Department, University of
  California, Berkeley, Feb 2009.

\bibitem{Naray-wmcsa00}
D.~Narayanan, J.~Flinn, and M.~Satyanarayanan.
\newblock Using history to improve mobile application adaptation.
\newblock In {\em Proc. of the 3rd IEEE Workshop on Mobile Computing Systems
  and Applications}, 2000.

\bibitem{Gurun-mobisys04}
S.~Gurun, C.~Krintz, and R.~Wolski.
\newblock Nwslite: A light-weight prediction utility for mobile devices.
\newblock In {\em Proc. of International Conference on Mobile Systems,
  Applications, and Services}, 2004.

\bibitem{adaptive}
Xiaohui Gu, Klara Nahrstedt, Alan Messer, Ira Greenberg, and Dejan Milojicic.
\newblock Adaptive offloading inference for delivering applications in
  pervasive computing environments.
\newblock In {\em PERCOM '03: Proceedings of the First IEEE International
  Conference on Pervasive Computing and Communications}. IEEE Computer Society,
  2003.

\bibitem{Hunt99-coign}
Galen~C. Hunt, Michael~L. Scott, Galen~C. Hunt, and Michael~L. Scott.
\newblock The coign automatic distributed partitioning system.
\newblock In {\em Proc. of the 3rd Symposium on Operating Systems Design and
  Implementation}, pages 187--200, 1999.

\bibitem{R-OSGi}
J.~S. Rellermeyer, G.~Alonso, and T.~Roscoe.
\newblock R-osgi: distributed applications through software modularization.
\newblock In {\em Proc. of the ACM/IFIP/USENIX International Conference on
  Middleware}, 2007.

\bibitem{cloudlet1}
Mahadev Satyanarayanan, Paramvir Bahl, Ramon Caceres, and Nigel Davies.
\newblock The case for vm-based cloudlets in mobile computing.
\newblock {\em IEEE Pervasive Computing}, 2009.

\bibitem{cloudlet2}
Adam Wolbach, Jan Harkes, Srinivas Chellappa, and M.~Satyanarayanan.
\newblock Transient customization of mobile computing infrastructure.
\newblock In {\em MobiVirt '08: Proceedings of the First Workshop on
  Virtualization in Mobile Computing}. ACM, 2008.

\bibitem{paranoid}
Georgios Portokalidis, Philip Homburg, Kostas Anagnostakis, and Herbert Bos.
\newblock Paranoid android: Versatile protection for smartphones.
\newblock In {\em Proceedings of the 26th {Annual Computer Security
  Applications Conference (ACSAC)}}, Austin, Texas, December 2010.

\bibitem{wowmom}
Eric~Y. Chen and Mistutaka Itoh.
\newblock Virtual smartphone over {IP}.
\newblock In {\em Proceedings of the IEEE International Symposium on A World of
  Wireless, Mobile and Multimedia Networks (WoWMoM)}, pages 1--6, Los Alamitos,
  CA, USA, June 2010. IEEE Computer Society.

\bibitem{xen}
Paul Barham, Boris Dragovic, Keir Fraser, Steven Hand, Tim Harris, Alex Ho,
  Rolf Neugebauer, Ian Pratt, and Andrew Warfield.
\newblock Xen and the art of virtualization.
\newblock In {\em SOSP '03: Proceedings of the nineteenth ACM symposium on
  Operating systems principles}, pages 164--177, New York, NY, USA, 2003. ACM.

\bibitem{qemu}
Fabrice Bellard.
\newblock Qemu, a fast and portable dynamic translator.
\newblock In {\em Proceedings of the annual conference on USENIX Annual
  Technical Conference}, ATEC '05, pages 41--41, Berkeley, CA, USA, 2005.
  USENIX Association.

\bibitem{powertutor}
L.~Zhang, B.~Tiwana, Z.~Qian, Z.~Wang, R.~P. Dick, Z.~Mao, and L.~Yang.
\newblock Accurate online power estimation and automatic battery behavior based
  power model generation for smartphones.
\newblock In {\em Proc. Int. Conf. Hardware/Software Codesign and System
  Synthesis}, 2010.

\bibitem{umts}
H.~Holma and A.~Toskala.
\newblock {\em HSDPA/HSUPA for UMTS: High Speed Radio Access for Mobile
  Communications}.
\newblock John Wiley and Sons, 2006.

\bibitem{3g_radio}
Feng Qian, Zhaoguang Wang, Alexandre Gerber, Zhuoqing~Morley Mao, Subhabrata
  Sen, and Oliver Spatscheck.
\newblock Characterizing radio resource allocation for 3g networks.
\newblock In {\em Proceedings of the 10th annual conference on Internet
  measurement}, IMC '10, pages 137--150, New York, NY, USA, 2010. ACM.

\bibitem{game}
{The Computer Language Benchmark Game} http://shootout.alioth.debian.org/.

\end{thebibliography}
